\documentclass[11pt]{article}

\usepackage{color}

\definecolor{indigo}{RGB}{0,0,120}
\usepackage[colorlinks=true, linkcolor=indigo, citecolor=blue, urlcolor=indigo, backref=page]{hyperref}

\usepackage{amsmath,amssymb} 

\usepackage[pdftex]{graphicx}

\usepackage{calligra}

\usepackage[skip=3pt,labelfont=normal,labelsep=colon]{caption}
\usepackage[skip=3pt]{subcaption} 
\usepackage{wrapfig}

\usepackage{cancel}
\usepackage[normalem]{ulem}
\usepackage{bm}

\newcommand{\tl}[1]{\tilde{#1}}

\newcommand{\dodo}[2]{\frac {\partial #1}{\partial #2}}

\def\ZZZ{{\hbox{ Z\kern-1.6mm Z}}}
\def\RRR{{\hbox{ R\kern-2.4mm R}}}
\def\CCC{{\hbox{ C\kern-2.0mm C}}}
\def\zzz{{\hbox{z\kern-1mm z}}}

\def\ZZZ{{\mathbb Z} }
\def\RRR{{\mathbb R} }
\def\CCC{{\mathbb C} }

\def\({\left(}
\def\){\right)}
\def\[{\left[}
\def\]{\right]}

\newcommand{\qeq}{{\hbox{=\kern-2.3mm ? \kern.5mm }}}
\renewcommand{\qeq}{=}

\newcommand{\eps}{\epsilon}

\newcommand{\ra}{\rangle}
\newcommand{\la}{\langle}

\usepackage{lipsum} 
\usepackage[]{mdframed}

\newcommand{\framedtext}[1]{%
\par%
\noindent\fbox{%
    \parbox{\dimexpr\linewidth-2\fboxsep-2\fboxrule}{#1}%
}%
}

\newcommand{\al}{\alpha}

\def\cl0{\tilde c_0}

\def\one{{\hbox{ 1\kern-.8mm l}}}
\def\zero{{\hbox{ 0\kern-1.5mm 0}}}

\def\ovq{\overline{q}}
\def\ovr{\overline{r}}

\def\ovk{\overline{k}}

\def\sig{\sigma}
\def\tht{\theta}
\def\del{\delta}

\begin{document}

\title{
	Smoothness of Classical Limit in KMOC Formalism
\author{{\sc Pritish Sinha}
\\ \small
Chennai Mathematical Institute, H1, SIPCOT IT Park, Siruseri, Kelambakkam 603103, India\\
\\ \small
Email: {\tt pritish@cmi.ac.in}}
\date{June 21, 2026 
\vspace{0.1cm}
}}
	\maketitle
\abstract{ In this paper, we revisit the smoothness of the classical limit of inclusive observables in the formalism developed by Kosower, Maybee and O'Connell (KMOC). Building on the earlier work \cite{KMOC, exp-S-Mat, exp-S-Mat-KMOC}, we prove that the classical limit of three classes of inclusive observables, namely scattering angle, radiative field and angular impulse is smooth and does not suffer from any so-called superclassical divergences at all orders in perturbation. We use scalar QED as a model. Our analysis goes some way in showing that KMOC formalism can be used to compute classical radiation by simply focusing on all the terms that scale as $\hbar^{0}$, as all the terms that scale with inverse power of $\hbar$ vanish. Through this analysis, we have successfully isolated the classes of terms that contribute to the classical limit, thereby allowing direct and more efficient computations of physical observables.}

{\footnotesize \tableofcontents}

\small
	
	\section{Introduction}
Kosower, Maybee and O'Connell have developed a formalism to obtain classical observables using the on-shell S-matrix \cite{KMOC}. The essential idea is based on the fundamental realization that in classical scattering, measurable quantities in the future can be realized as a classical limit of a certain class of inclusive observables which are determined by a second-order polynomial of the scattering amplitude \cite{mizera-hoffie-caronhuot}. The power of this formalism lies in its universality in computing various classical observables. For example, KMOC has been applied to calculate the scattering angle, the radiated momentum \cite{c:Grab-Brem-Rev-Unt,c:rad-clas-grav-obs-scat-amp} and radiated angular momentum \cite{c:Rad-Ang-Mom-Dis-eff} at 2-loops $O(G^{3})$ for the scattering of spinless black holes.

Thanks to the unitarity of the S-matrix, observables susceptible to the radiation reaction (such as the scattering angle at $O(G^{3})$ in gravity) are cleanest to compute via KMOC methods. This is in contrast to a direct classical computation where radiation reaction effects have to be accounted for separately by using additional terms in the equations of motion such as Abraham-Lorentz-Dirac (ALD) acceleration terms. The formalism has been extended to compute impulse and radiation for the scattering of spinning particles and black holes at $1PM$ order \cite{c:obs-amp-spin-part-bh,c:BH-scat-spin-dir-min-coup-amp,c:kerrBH-elem-part,c:class-obs-exp-spin-fact,c:class-obs-coh-spin-amp}. Computation of gravitational (and electromagnetic) waveforms and the inclusion of massless bosonic particles in initial states via KMOC have been considered in \cite{c:wv-from-amp}. The sub-leading $PM$ waveform has been obtained in \cite{mizera-hoffie-caronhuot,c:sub-lead-scat-wav-form, c:Rad-rxn-1loop, c:In-exp-clas-grav-scat, c:Ap2-In-exp-clas-grav-scat, c:1loop-gtav-brem-wav-heft} by taking the classical limit of a 1-loop 5-pt amplitude with one graviton emission using the KMOC formalism. This has been extended to compute the waveforms for the scattering of spinning particles \cite{c:spin-wav-KMOC-LO,c:sum-wav-5pt-amp}. Studying the classical limit of thermal currents \cite{c:Scat-amp-therm-loop} and deriving classical soft theorems \cite{subnD=4,KMOC-soft} from amplitudes are a few other areas where KMOC has found applications. KMOC has also been extended to compute classical observables in curved spacetimes \cite{c:class-phy-curv-back} and classical Yang-Mills observables \cite{c:clas-yang-mill-obs-amp}. KMOC has also been used to derive constraints on amplitudes itself by demanding consistency with its classical limit, such as negligible variance \cite{c:un-pr-class-amp} or classical soft theorems \cite{sof-cons-KMOC}.

In a nutshell, KMOC formalism is applicable when we scatter massive classical particles (with potentially infinitely many multipole moments) at a large impact parameter \cite{KMOC} or when we perturb a classical particle with a gravitational or electromagnetic wave \cite{c:wv-from-amp}. By giving a precise algorithm to obtain the classical limit at the level of loop integrands inside the amplitude, KMOC formalism has become one of the most indispensable tools in the efficient computation of classical radiative observables using all the modern tools of computing scattering amplitudes. More in detail, the classical limit is taken by scaling all the momenta (real and virtual) associated with massless particles to zero at the level of the loop integrand. A price one inevitably pays in this approach is the occurrence of the so-called superclassical terms at the intermediate steps of the computation which scale inversely with $\hbar$ and potentially make the classical limit divergent. However, in all computations performed so far using the KMOC formalism, one observes that the superclassical terms mutually cancel each other.\footnote{For scattering angle, such terms arise at next to leading order (NLO) in the coupling and their cancellation mimics the cancellation of infra-red divergence in inclusive cross section.} Nevertheless, a systematic proof that the KMOC classical limit is smooth for all observables would be desirable. 

Before KMOC, such divergences had appeared in the background field QED, such as in the computation of radiation reaction while taking the classical limit of photon emission due to particles accelerated by the plane wave field. The divergences were shown to cancel at one loop in \cite{c:rad-rxn-str-fld} and at all loops in \cite{c:res-of-quan-rad-rxn-pln-wave}. The worldline formalism is another method to obtain classical observables where superclassical terms are avoided \cite{c:PM-EFT-CBD, c:CBH-WQFT}. The relation between the Worldine formalism and the KMOC formalism has also been established \cite{c:KMOC-WQFT, c:WL-Sa}.

In another direction, the exponential representation of the S-matrix has been shown to be particularly useful in providing a smooth semi-classical limit of the scattering using velocity cuts \cite{exp-S-Mat}. It is used to capture the classical dynamics for few leading orders in coupling, which includes the conservative sector and parts of the radiative sector, but the procedure lacks generalization. This has been addressed by combining the KMOC formalism with the exponential representation of the S-matrix in \cite{exp-S-Mat-KMOC}. This has also ensured that the formula for obtaining the classical scattering angle using the amplitude is smooth for a certain class of terms. In particular, the authors used the exponential representation of the S-matrix inside the KMOC formalism and showed that there are no superclassical terms at all orders in the perturbation theory for those terms. This is achieved by exploiting the smoothness of the semi-classical limit of $\log\hat{S}$ in the exponential formalism. It leads to compact formulas for classical observables incorporating all conservative and radiative contributions. The classical limit is expanded in powers of $\log\hat{S}$ as opposed to the perturbation expansion. This makes superclassical terms far more efficient to handle and leads to a very elegant representation of the KMOC formalism. This expansion is then used to organize the classical contributions on the basis of intermediate states inserted between $\log\hat{S}$ operators and it is shown that the classical limit of the scattering angle for the intermediate massive $2$-particle states is always smooth. These intermediate massive $2$-particle states are referred to as elastic channels.

In the current work, we generalize the proof to include the inelastic channels as well, thus completing the proof for the smoothness of the classical limit. We use the scalar QED as the model, the same as in the main KMOC paper \cite{KMOC}, but the results are extendable to Gravity. We have discussed this in the Discussion section. We are further motivated by the following question: Can the exponential representation of the S-matrix be used to prove that \emph{any} observable computed via KMOC formalism admits a smooth classical limit? Although we do not answer this question in full generality, we build on the results in \cite{exp-S-Mat-KMOC} to make further progress on this question. More in detail, we show that for the case of electromagnetic scattering of the scalar particles, KMOC formulae admit a smooth classical limit for the radiative electromagnetic field and angular impulse of the scalar particles to all orders in the perturbative expansion. The reason we analyse these two observables is inspired by the primary question that we want to answer. The radiative field is clearly in a different category of observable than scattering angle as the former pertains to a computation in the conservative sector while the latter measures the loss of matter momenta, which are radiated away to null infinity. The angular impulse, while closer to the scattering angle in the sense that it is non-zero even within the conservative sector, is qualitatively different from the latter since the asymptotic states are not eigen states of an angular momentum operator. Thus, we believe that the proof of a well-defined classical limit for these two observables adds to the body of evidence that the KMOC classical limit is smooth. 

 Furthermore, a systematic understanding of the cancellation mechanism, as presented in this work, allows for more streamlined computations of physical observables. In this paper, we have shown that the classical limits contain only those terms that have elastic channels or inelastic channels with only single-photon insertions between $\log\hat{S}$ operators. Any higher number of photon insertions between them is shown to be necessarily quantum. Thus, we have isolated the set of terms that contribute to the classical limit. We have provided explicit recursive expressions for elastic channels, which remain after the cancellation of superclassical terms. For inelastic channels, the proof allows us to directly pick out the terms in the amplitude that scale classically. Moreover, the understanding of the proof also let us isolate the physical ingredients that are crucial for the cancellation of superclassical terms and may further extend the proof to a wider class of observables and interactions. We have outlined the possibilities in the Discussion.

\section{KMOC formalism in Exponential Representation}
	
    The exponential representation of the S-matrix \cite{exp-S-Mat} is given by:
	\begin{align}
	S = e^{i N / \hbar}.
	\end{align} 
    As argued in \cite{exp-S-Mat, exp-S-Mat-KMOC}, because the $N$-operator is already in the exponent of the ${S}$-matrix, the matrix elements of $N$, by construction, are free from superclassical terms. This means that the matrix elements of $N$ have a smooth semi-classical expansion in $\hbar$. This implies that when the classical limit of $\la a |N | b\ra$ is taken using the KMOC formalism, it scales as $\hbar^k$, where $k$ is fixed for any order in the coupling constant for a given incoming state $\la a |$ and outgoing state $| b \ra$. We will denote it by $\la a |N | b\ra \sim \hbar^k$ from now on. Since tree-level diagrams always contribute classically \cite{c:clas-phys-quan-loop}, $k$ can be determined for each matrix element of $N$ by calculating the corresponding tree-level diagrams and taking the classical limit. KMOC formalism provides us with an algorithm for computing asymptotic classical observables for the large impact parameter scattering of two particles satisfying $|b| >> l_w >> l_c $, where $b$ is the impact parameter, $l_w$ is the spread of the wavepacket, and $l_c$ is the Compton wavelength. The asymptotic observables in which we are interested are given by:
\begin{align}
 \Delta O = S^\dag O S - O= e^{- i N / \hbar} \, O \, e^{ i N / \hbar} = \sum_{n \geq 1} \frac{(-i)^n}{\hbar^n n!} [N,[N,\ldots[N,O]]],
\label{eq:D-O-N_opt}
\end{align} 
in the exponential formalism \cite{exp-S-Mat-KMOC}.
Given $p_1$ and $p_2$ as the momenta of incoming particles, the classical limit of the asymptotic observable \cite{exp-S-Mat-KMOC} is given by: 
\begin{align}
\la \Delta O \ra (p_1, p_2, b) = \lim_{\hbar \to 0}  \int \hat{d}^4 p_1' \hat{d}^4 p_2' \: \hat{\del}(p_1'^2 - m_1^2) \hat{\del}(p_2'^2 - m_2^2)  \: e^{-ib \cdot \ovq_1} \, \la p_1', p_2' |\Delta O| p_1, p_2 \ra,
\label{eq:clas-exp-the-or}
\end{align}
where $p_1' = p_1 + \hbar \ovq_1$, $p_2' = p_2 + \hbar \ovq_2$.
The hat notation is defined as: $\hat{\del}(x) = 2\pi \del(x)$ and $\hat{d}x = dx/(2\pi)$. In cases where $ \la p_1', p_2' | \Delta O | p_1, p_2 \ra $ can be factorized as:
\begin{align}  
 \la p_1', p_2' | \Delta O | p_1, p_2 \ra = \Delta O (p_1, p_2 \to p_1', p_2') \hat{\del}^{(4)}(p_1'+p_2'-p_1-p_2),
\label{eq:defo}
\end{align}
 the $\la \Delta O \ra (p_1, p_2, b)$ is given by:
\begin{align}
\la \Delta O \ra (p_1, p_2, b) = \lim_{\hbar \to 0} \hbar^2  \int \hat{d}^4 \ovq \: \hat{\del}(2 p_1 \cdot \ovq) \hat{\del}(2 p_2 \cdot \ovq) e^{-ib \cdot \ovq}  \:\overline{\Delta O }(p_1, p_2 \to p_1+ \hbar \ovq, p_2-\hbar \ovq),
\label{eq:clas-exp}
\end{align}
where $p_1' = p_1 + \hbar \ovq$ and $p_2' = p_2 - \hbar \ovq$.
We place a bar on the function after taking the classical limit. For taking the limit, loop momenta, exchange momenta and radiated massless momenta are scaled linearly with $\hbar$. In other words, $q$ is replaced by $\hbar \ovq$, where $\ovq$ is the wavenumber that is considered finite and $\hbar$ is scaled to 0. Our aim in this paper is to prove that the asymptotic observable $\la \Delta O \ra (p_1, p_2, b)$, as defined by Eqn.(\ref{eq:clas-exp-the-or}), is free of superclassical terms. In other words, $\la \Delta O \ra (p_1, p_2, b) \sim \hbar^0$. To do this, it is useful to define the following \cite{exp-S-Mat-KMOC}:
\begin{align}
A_n^O = \frac{1}{\hbar^n} \underbrace{[N,[N, \ldots,[N,O]]]}_{n\:\text{times}}.
\label{eq:anono}
\end{align}
Eqn.(\ref{eq:anono}) can be written in the following recursive form:
\begin{align}
A_n^O = \frac{1}{\hbar}[N, A_{n-1}^O] = A_1^{A_{n-1}^O}. 
\label{eq:recur}
\end{align} 
Substituting Eqn.(\ref{eq:anono}) into Eqn.(\ref{eq:D-O-N_opt}), we get:
\begin{align}
\Delta O =  \sum_{n \geq 1} \frac{(-i)^n}{ n!}A_n^O.
\label{eq:o-A-n}
\end{align}
So, showing $A_n^O$ to be free of superclassical terms for an arbitrary $n$ is sufficient to establish the classical limit of $\Delta O$ is free of  superclassical terms.
Since the matrix elements of $A_n^O$ depend on the matrix elements of $N$, we need to determine $N$. This can be done by equating the exponential representation of the S-Matrix with its Born representation:
\begin{align}
I + i\frac{T}{\hbar} = e^{iN/\hbar}.
\label{eq:exp-born}
\end{align} 
We perturbatively expand $N$ and $T$  as:
\begin{align}
N = e^2 N_2 + e^3 N_3 + e^4 N_4 + \ldots,
\end{align}
\begin{align}
T = e^2 T_2 + e^3 T_3 + e^4 T_4 + \ldots.
\end{align}
 Substituting them into Eqn.(\ref{eq:exp-born}), we can write the matrix elements of $N$ in terms of amplitudes perturbatively:
\begin{align}
& N_2 = T_2, 
\nonumber\\
& N_3 = T_3, 
\nonumber\\
& N_4 = T_ 4 - \frac{i}{2 \hbar} T_2^2, 
\nonumber\\
& N_5 = T_ 5 - \frac{i}{2 \hbar} (T_2 T_3 + T_3 T_2) \dots.
\label{eq:N-T-reln}
\end{align}

$\la b| (T_k/\hbar) | a \ra$ is simply the amplitude at order $o(e^k)$, which can be calculated using the Feynman rules. Thus, we can determine how $\la b| N | a \ra$ scales in the semi-classical limit by knowing how the corresponding tree-level diagram scales. This scaling, then, ensures cancellation of superclassical terms for higher orders of coupling constant. It implies an intricate cancellation of superclassical terms among themselves that appear due to both loop diagrams and cut diagrams.

\section{Classical Limit of Linear Impulse}
 
In this section, we show that the classical limit of the linear impulse is smooth in the KMOC formalism. In fact, we consider an observable $\hat{O}$ that can be any polynomial in linear momentum operators $\hat{P}_{1}, \hat{P}_{2}$, defined by:
\begin{align}
\hat{O}\,|p_1, p_2, r_1, \sig_1, r_2, \sig_2. \ldots r_k, \sig_k \ra = O(p_1, p_2)\,|p_1, p_2, r_1, \sig_1, r_2, \sig_2, \ldots r_k, \sig_k \ra,
\label{eq:O(p1p2)p1p2}
\end{align}
where $|p_1, p_2, r_1, \sig_1, r_2, \sig_2, \ldots r_k, \sig_k \ra$ represents the momenta of 2 massive scalars ($p_1, p_2$) and $k$ photons ($r_1, \cdots ,r_k$) with corresponding photon helicities ($\sigma_1, \cdots ,\sigma_k$). 
For such an observable, we prove Eqn.(\ref{eq:clas-exp}) to be superclassical free. Substituting Eqn.(\ref{eq:o-A-n}) into Eqn.(\ref{eq:clas-exp}), and demanding that it be free of superclassical terms, it suffices to prove that Eqn.(\ref{eq:defo}) is superclassical free for $A_n^O$, i.e.,
\framedtext{
\begin{align}
  \la p_1', p_2' | A_n^O| p_1, p_2 \ra  \sim \frac{1}{\hbar^2} \, \hat{\del}^{(4)} (p_1'+p_2'-p_1-p_2),
 \label{eq:main-stat}
\end{align}
for any arbitrary $n$ in the classical limit.
}
\\
\\
For simplicity, we represent photon states as:
\begin{align}
|\Upsilon_k^{\vec{r},\vec{\sig}}\ra = |r_1, \sig_1, r_2, \sig_2. \ldots r_k, \sig_k \ra.
\label{eq:new-sym}
\end{align}
We use Eqn.(\ref{eq:O(p1p2)p1p2}) to evaluate $ \la p_1', p_2', \Upsilon_k^{\vec{r},\vec{\sig}} | A_1^O |  p_1, p_2 \ra $, where $p_1'= p_1 + q$ and $p_2' = p_2 -q-\sum_i r_i$. Upon evaluation, we find:
\begin{align}
\frac{1}{\hbar}  \la p_1', p_2', \Upsilon_k^{\vec{r},\vec{\sig}} |[N, O]| p_1, p_2 \ra  = \frac{1}{\hbar} \la p_1', p_2', \Upsilon_k^{\vec{r},\vec{\sig}} |N| p_1, p_2 \ra (O(p_1,p_2) - O(p_1',p_2')).
\label{eq:Noon}
\end{align}
Factoring the delta function from the matrix element of $N$:
\begin{align}
\la p_1', p_2', \Upsilon_k^{\vec{r},\vec{\sig}} | N |p_1, p_2  \ra  = \, N ( p_1, p_2  \to p_1', p_2', \Upsilon_k^{\vec{r},\vec{\sig}} ) \, \hat{\del}^{(4)}(\tl{p}_1 + \tl{p}_2 - \sum_i r_i - p_1' - p_2' ).
\label{eq:del-fact}
\end{align}
 Expanding $O(p_1',p_2')$ semi-classically:
\begin{align}
O(p_1',p_2') = O(p_1,p_2) + \hbar \left(  \ovq^\mu \nabla_\mu -  \sum_i \ovr^\mu_i \dodo{}{p_2^\mu} \right)O(p_1,p_2),
\label{eq:O-expnd}
\end{align}
where
\begin{align}
\nabla_\mu = \dodo{}{p_1^\mu} - \dodo{}{p_2^\mu}.
\label{eq:nabla}
\end{align}
We insert Eqn.(\ref{eq:del-fact}) and Eqn.(\ref{eq:O-expnd}) into Eqn.(\ref{eq:Noon}) to obtain:
\begin{align}
 \la & p_1', p_2',  \Upsilon_k^{\vec{r},\vec{\sig}} |A_1^O | p_1, p_2 \ra  =
\nonumber\\
 &  - N ( p_1, p_2 \to p_1', p_2', \Upsilon_k^{\vec{r},\vec{\sig}} )\, \hat{\del}^{(4)}(p_1 + p_2 - \sum_i r_i - p_1' - p_2' )
\left(  \ovq^\mu \nabla_\mu -  \sum_i \ovr^\mu_i \dodo{}{p_2^\mu} \right) O(p_1,p_2).
\label{eq:A1phot}
\end{align}

Using Eqn.(\ref{eq:N-T-reln}), we can write $N(p_1, p_2 \to p_1', p_2')$ in terms of an amplitude.
For $k = 0$, the tree level $N(p_1, p_2 \to p_1', p_2')$ in the scalar QED is given by: 
\begin{align}
\lim_{\hbar \to 0} N(p_1, p_2 \to p_1', p_2') = \hbar  A(p_1, p_2 \to p_1', p_2') = Q_1Q_2\frac{\overline{e}^2}{\hbar^2} \frac{4p_1 \cdot p_2}{\ovq^2}.
\label{eq:N2to2-1/h2}
\end{align}
Thus, substituting Eqn.(\ref{eq:N2to2-1/h2}) into Eqn.(\ref{eq:A1phot}), we find $ \la p_1', p_2' | A_1^O | p_1, p_2 \ra \sim  \hbar^{-2} \, \hat{\del}^{(4)} (p_1'+p_2'-p_1-p_2)
$. Thus, it is superclassical free. Since the tree-level contribution is sufficient to determine how $\la p_1', p_2' |N | p_1, p_2 \ra $ scales with $\hbar$ for any order in $e$. This ensures that all superclassical terms that appear in higher-order $o(e^s)$ $(s>2)$ diagrams must cancel among each other. 

Next, we need to prove that $ \la p_1', p_2' | A_n^O | p_1, p_2 \ra $ scales as $\sim \hbar^{-2}$ after factoring out the delta function for arbitrary $n$. We prove this by induction. We have already shown it for $n = 1$. We assume that it holds for $n-1$ in Eqn.(\ref{eq:recur}) and then prove it for $n$. Inserting intermediate states between $N$ and $A_{n-1}^O$:
\begin{align}
 \la p_1', p_2' | A_n^O | p_1,  p_2  \ra & =  \, \frac{1}{\hbar}\la p_1', p_2' | [N, A_{n-1}^O] | p_1, p_2 \ra   
\nonumber\\
 & =  \, \frac{1}{\hbar}\sum_{k \geq 0} \int d \phi(\tl{p}_1, \tl{p}_2, \Upsilon_k^{\vec{s},\vec{\rho}}) (\la p_1', p_2' | N| \tl{p}_1, \tl{p}_2,\Upsilon_k^{\vec{s},\vec{\rho}}\ra \, \la \tl{p}_1, \tl{p}_2,\Upsilon_k^{\vec{s},\vec{\rho}} | A_{n-1}^O| p_1, p_2 \ra  
\nonumber\\
& - \la p_1', p_2' | A_{n-1}^O | \tl{p}_1, \tl{p}_2,\Upsilon_k^{\vec{s},\vec{\rho}} \ra \, \la \tl{p}_1, \tl{p}_2,\Upsilon_k^{\vec{s},\vec{\rho}} | N | p_1, p_2 \ra ),
\label{eq:chan}
\end{align}
where $k$ is the number of photons and the measure is defined as:
\begin{align}
d \phi(\tl{p}_1, \tl{p}_2, \Upsilon_k^{\vec{s},\vec{\rho}}) = \frac{1}{k!}\sum_{\vec{\rho}} d \phi(\tl{p}_1) d \phi(\tl{p}_2) \prod_{i=1}^{k} d \phi(s_i),
\end{align}
where $d \phi(\tl{p}) = \hat{d}^{4}\tl{p}\,\hat{\delta}(\tl{p}^2 - m^2) = \hat{d}^{3}\tl{p}/2E_{\tl{p}}$ and $d \phi(\tl{s})  = \hat{d}^{4}\tl{s}\,\hat{\delta}(s^2) = \hat{d}^{3}\tl{s}/2s^0$. 
We refer to intermediate 2-particle states ($k=0$) as an elastic channel and intermediate states with any additional photon as an inelastic channel ($k>0$). Note that in the KMOC formalism, we work in the large impact parameter regime, thus, only those channels contribute that allow for small momentum deviations. This forbids massive particle production in inelastic channels, since 
\begin{align}
|\, p^0_1 + p^0_2 - p'^0_1 - p'^0_2 \, | = \hbar \, |\, \ovq^0_1+\ovq^0_2 \, |  <<  m.
\end{align}
Thus, no inelastic channel with massive particle production can contribute classically.

We divide the proof into two parts: first, we only consider terms containing elastic channels and show the cancellation of superclassical terms. This is the same proof as done in \cite{exp-S-Mat-KMOC} and shown in the Appendix \ref{ap:el-ch-cny}. Following this, we go straight to extend the proof to terms containing inelastic channels.

\subsection{Inelastic Channel Contributions}
\label{s:in-cc}

We consider the terms in Eqn.(\ref{eq:chan}), where we insert an inelastic channel corresponding to some fixed $k \geq 1$ number of photons for $n \geq 2$. For smoothness of the classical limit, we need to show that the $1/\hbar$ factor in Eqn.(\ref{eq:chan}) gets canceled in the classical limit. We prove this using induction, where we assume that $A_{n-1}^O$ is superclassical free. This means that all $1/\hbar$ factors in Eqn.(\ref{eq:anono}) is assumed to get canceled in case of $A_{n-1}^O$ in the classical limit. For $A_1^O$, we have already shown in Eqn.(\ref{eq:A1phot}) that $1/\hbar$ cancels for any of its matrix elements. Next, we prove the same for $A_n^O$ and show the following for inelastic contributions in Eqn.(\ref{eq:chan}):
\\
\framedtext{
\begin{align}
 \frac{1}{\hbar}\int d \phi(\tl{p}_1, \tl{p}_2, & \Upsilon_k^{\vec{s},\vec{\rho}}) (\la p_1', p_2' | N| \tl{p}_1, \tl{p}_2,\Upsilon_k^{\vec{s},\vec{\rho}}\ra \la \tl{p}_1, \tl{p}_2,\Upsilon_k^{\vec{s},\vec{\rho}} | A_{n-1}^O| p_1, p_2 \ra  
\nonumber\\
- & \la p_1', p_2' | A_{n-1}^O | \tl{p}_1, \tl{p}_2,\Upsilon_k^{\vec{s},\vec{\rho}} \ra \la \tl{p}_1, \tl{p}_2,\Upsilon_k^{\vec{s},\vec{\rho}} | N | p_1, p_2 \ra )
 \nonumber\\
 & \sim \hbar^{k-3} \, \hat{\del}^{(4)} (p_1'+p_2'-p_1-p_2) \quad \text{for} \quad k\geq 1,
 \label{eq:3pt1pf}
\end{align}
where $n \geq 2$. Thus, it also implies that the only classical contribution can come from the inelastic channel consisting of one photon ($k=1$). For any $k>1$ number of photons, the contribution is necessarily quantum. 
}
\\
\\
In this section, we restrict ourselves to photon insertions having non-zero momenta. A separate calculation for photons having trivial momentum, also known as static mode photons, is carried out in the next section \ref{0-freq-phot}.
\\
The order of $\hbar^s$  in $\la p_1', p_2', \Upsilon_n^{\vec{r},\vec{\sig}}| N| p_1, p_2 \ra  $ can be determined by the tree-level diagram.
The tree-level contribution to $p_1, p_2 \to p_1', p_2', \Upsilon_n^{\vec{r},\vec{\sig}}$ is given by: 
\begin{align}
  \la p_1', p_2', \Upsilon_n^{\vec{r},\vec{\sig}}| N| p_1, p_2 \ra   = & \, \hbar \, \la p_1', p_2', \Upsilon_n^{\vec{r},\vec{\sig}} | \frac{T}{\hbar}| p_1, p_2 \ra   
\nonumber\\
= &\, \hbar\, \textit{A}(p_1, p_2 \to p_1', p_2', \Upsilon_n^{\vec{r},\vec{\sig}}) \:\hat{\delta}^{(4)}(p_1'+ p_2'+ r_1 +\ldots r_n - p_1 - p_2),
\label{eq:N-intrm-T}
\end{align}
where $p_1'= p_1 + q$ and $p_2' = p_2 -q-\sum_i r_i$. 

To determine $o(\hbar^s)$ for $N(p_1, p_2 \to p_1', p_2', \Upsilon_n^{\vec{r},\vec{\sig}})$, we need to calculate tree-level diagrams and then take the classical limit using the KMOC prescription. Since the diagram contains $n$ external photons that can be emitted from either of the two charged massive scalars, $(m_1, Q_1)$ and $(m_2, Q_2)$, there are exactly $2^n$ distinct ways to distribute the photons between the two scalars. 

\begin{figure}
	\centering
	\includegraphics[width=200pt]{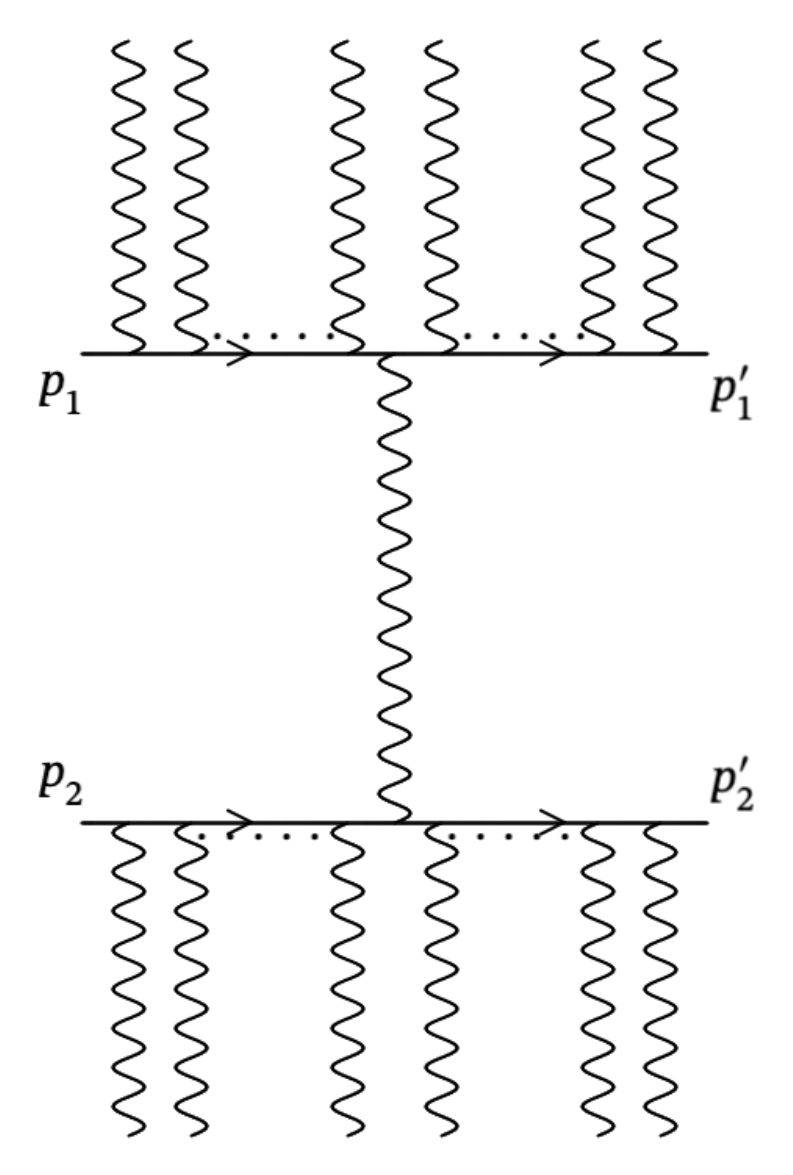}
	\caption{Tree-level diagram with only 3 pt. vertices for $p_1, p_2 \to p_1', p_2', \Upsilon_n^{\vec{r},\vec{\sig}}$ amplitude}
	\label{fig:fey-n-phot}
\end{figure}

We consider contributions only from diagrams consisting of 3-vertices as any contribution from a 4-point vertex is necessarily sub-leading. This is because a 4-point vertex effectively replaces two 3-point vertices and an intermediate massive propagator. In scalar QED, the product of two 3-point vertices and the intermediate propagator, given by $(ieQ)^2(i/p\cdot k)$, scales as $1/\hbar^2$. Since the 4-point vertex, given by $2ie^2Q^2g^{\mu\nu}$, scales as $1/\hbar$, this makes a 4-point vertex contribution sub-leading. Thus, the diagrams contributing to the leading order must only contain $3$-point vertices, as depicted in Fig. (\ref{fig:fey-n-phot}). 

Consider an arbitrary such diagram with $j$ photons with corresponding momenta $r_{i_1}, r_{i_2} \ldots r_{i_j}$ that get emitted from mass $(m_1, Q_1)$ and remaining $n-j$ photons with momenta $r_{i_{j+1}}, r_{i_{j+2}} \ldots r_{i_n}$ that get emitted from $(m_2, Q_2)$. 

To compute this, we need to break down the computation into pieces. We consider an incoming scalar $(m_1, Q_1)$ emitting photons with momenta $r_{i_1}, r_{i_2} \ldots r_{i_k}$. These photons can be permuted in $k!$ ways, each of which contributes to amplitude. Consider a permutation $\pi \in S_k$, for which the momenta of the scalar lines are given by: 
\begin{align}
p_1, \, p_1 - r_{i_{\pi(1)}},\, p_1 - r_{i_{\pi(1)}} - r_{i_{\pi(2)}}, \ldots , p_1 -  \sum_{l = 1}^{k}r_{i_{\pi(l)}}. 
\end{align}
The on-shell condition for incoming massive particle, $p_1^2 = m_1^2$, implies: 
 \begin{align}
\left( p_1 -  \sum _{l = 1}^{k} r_{i_{\pi(l)}} \right)^2 - m_1^2 =  - 2 p_1 \cdot \left( \sum _{l = 1}^{k}r_{i_{\pi(l)}} \right) + \left(\sum _{l = 1}^{k} r_{i_{\pi(l)}} \right)^2.
 \end{align}
 In the classical limit, the quadratic terms are sub-leading in $\hbar$ and thus do not contribute. The factors contributed by the vertices along with their corresponding outgoing photon polarizations are given by: 
 \begin{align}
 (ie Q_1)(2 p_1 - r_{i_{\pi(1)}})\cdot\eps_p (r_{i_{\pi(1)}}), \,(ieQ_1)(2 p_1 - 2 r_{i_{\pi(1)}} - r_{i_{\pi(2)}})\cdot\eps_p (r_{i_{\pi(2)}}), \ldots, 
 \nonumber\\
 (ie Q_1)( 2 p_1 - 2 \sum_{l=1}^{k-1} r_{i_{\pi(l)}} -  r_{i_{\pi(k)}} ) \cdot \eps_p (r_{i_{\pi(k)}}).
 \end{align}
 $\eps_p (r_i)$ is the polarization vector $\sig_i$ of the $i^{th}$ photon. It is a shorthand notation for $\eps_{p(i)} (r_i)$, where $p(i)$ labels the polarization of the $i^{th}$ photon and $r_i$ is its corresponding momentum.
 Using the KMOC prescription, only the factors of $2 p_1 \cdot \eps_p(r_{i_{\pi(l)}})$ survive in the leading term. Thus, we can sum over the permutation to get the following contribution to the amplitude:
\begin{align}
\sum_ \pi \prod_{j=1}^{k} \frac{-i}{\hbar^{3/2}} \frac{(i\, \bar{e}\, Q_1) \, p_1 \cdot \eps_p (\ovr_{i_{\pi(j)}})}{ p_1 \cdot ( \sum _{l = 1}^{j}\ovr_{i_{\pi(l)}} ) - i \eps} \times \text{other parts of diagram}.  
\label{eq:up-hlf-fyn}
\end{align}
The above expression can be simplified using the following identity:
\begin{align}
\sum_\pi \prod_{j=1}^{k}  \frac{1}{ p_1 \cdot ( \sum _{l = 1}^{ j}\ovr_{i_{\pi(l)}} )} = \prod_{l = 1}^{k}\frac{1}{ p_1 \cdot \ovr_{i_{l}}}.
\label{eq:imp-id}
\end{align}
 The proof of the identity is shown in the Appendix \ref{ap:imp-id}.
Since  
\begin{align}
\prod_{j=1}^{k} p_1 \cdot \eps_p (\ovr_{i_{\pi(j)}}) = \prod_{j=1}^{k} p_1 \cdot \eps_p (\ovr_{i_j}), 
\label{eq:perm-inv}
\end{align}
which is invariant of any permutation $\pi$, substituting Eqn.(\ref{eq:imp-id}) into expression in (\ref{eq:up-hlf-fyn}), the expression becomes:
\begin{align}
 \frac{Q_1^k \: \bar{e}^k}{\hbar^{3k/2} } \: \prod_{l=1}^{k} \frac{p_1 \cdot \eps_p (\ovr_{i_l})}{ p_1 \cdot \ovr_{i_l}  - i \eps} \times \text{other parts of diagram}.  
\label{eq:perm-nop-12}
\end{align}
We can follow the same procedure as before for the outgoing scalar emitting photons. 
Next, we return to the arbitrary distribution of $j$ photons with momenta $r_{i_1}, r_{i_2} \ldots r_{i_j}$, emitted from the charged massive scalar $(m_1, Q_1)$ and the remaining $n-j$ photons with momenta $r_{i_{j+1}}, r_{i_{j+2}} \ldots r_{i_n}$, emitted from $(m_2, Q_2)$. Since, for a tree-level diagram, $j$ photons could be partitioned between $2$ sides of vertex i.e. incoming and outgoing scalar in $2^j$ ways, we define $\mathcal{P}_{j,2}$ as the collection of all such partitions that contribute to the amplitude. An internal photon line connects the 2 massive scalars. Let the partition be such that the internal photon attaches to a massive scalar $(m_1,\,Q_1)$ at a vertex with an incoming state that emits $k$ photons and an outgoing state that emits the remaining $j-k$ photons. 
Furthermore, consider the configuration of the external photon lines arranged in a permutation $\pi \in S_j$. The exchange momentum carried by the internal photon line in this case is: 
\begin{align}
\left( p_1' + \sum _{l = 1}^{k}r_{i_{\pi(l)}} \right) - \left(p_1 - \sum _{l = k+1}^{j}r_{i_{\pi(l)}}\right) = q + \sum_{l=1}^{j} r_{i_{\pi(l)}} = q + \sum_{l=1}^{j} r_{i_l}.
\label{eq:int-prop}
\end{align}
Hence, the momentum carried by the internal photon line does not depend on which permutation external photon lines are arranged (i.e., independent of $\pi$) nor does it depend on where the internal photon lines attach themselves (i.e., independent of $k$). The vertex factor, on the other hand, is given by $(p_1' + \sum _{l = 1}^{k}r_{i_{\pi(l)}}  + p_1 - \sum _{l = k+1}^{j}r_{i_{\pi(l)}})^\mu $, which, in the classical limit, simply reduces to $2 p_1^\mu$ for any permutation. 

So, including the internal photon propagator along with both incoming and outgoing scalars, we can write the contributions to the amplitude as:
\begin{align}
 \frac{Q_1^{j+1} \, \bar{e}^{j+1}}{\hbar^{(3j+5)/2} } \sum_{\{k\} \in \mathcal{P}_{j,2}}\prod_{l = 1}^{k} \frac{p_1 \cdot \eps_p (\ovr_{i_l})}{ p_1 \cdot \ovr_{i_l}  - i \eps}  \prod_{l = k+1}^{j} \frac{-p_1 \cdot \eps_p (\ovr_{i_l})}{ p_1 \cdot \ovr_{i_l}  + i \eps} \left( \frac{2 p_1^\mu g_{\mu \nu}}{\left(\ovq + \sum_{l=1}^{j} \ovr_{i_l}\right)^2 + i \eps} \right)
 \nonumber\\
 \times \text{other parts of diagram},
\end{align}
where we have summed over all partitions of $j$ photons onto both sides of the vertex. Since we sum over all $2^j$ partitions, the expression can be written as:
\begin{align}
 \frac{Q_1^{j+1} \, \bar{e}^{j+1}}{\hbar^{(3j+5)/2} } \,  \prod_{l = 1}^{j} p_1 \cdot \eps_p (\ovr_{i_l})\underbrace{\left( \frac{1}{ p_1 \cdot \ovr_{i_{l}} - i \eps}  - \frac{1}{p_1 \cdot \ovr_{i_{l}} + i \eps} \right)}_{i\,\hat{\del}\left(p_1 \cdot \ovr_{i_l}\right)} &\left( \frac{2 \, p_1^\mu g_{\mu \nu}}{\left(\ovq + \sum_{l=1}^{j} \ovr_{i_l}\right)^2 + i \eps} \right)
\nonumber\\
 &\times \text{other parts of diagram}.
\end{align}
Following the same analysis for the other scalar $(m_2, Q_2)$ with incoming momentum $p_2$ and outgoing momentum $p_2' = p_2 - q - \sum_{i=1}^{n} r_i$, the total amplitude at the tree level in the classical limit is given by:
\begin{align}
 \frac{i^n \, \bar{e}^{n+2}}{\hbar^{(3n/2)+3} } \sum_{\{j\} \in \mathcal{P}_{n,2}} & Q_1^{j+1} \, Q_2^{n-j+1} \prod_{l = 1}^{j} p_1 \cdot \eps_p (\ovr_{i_l}) \, \hat{\del}(p_1 \cdot \ovr_{i_l})\prod_{l = j+1}^{n} p_2 \cdot \eps_p (\ovr_{i_l}) \, \hat{\del}(p_2 \cdot \ovr_{i_l})
\nonumber\\
& \times \left( \frac{4p_1 \cdot p_2}{\left(\ovq + \sum_{l=1}^{j} \ovr_{i_l}\right)^2 + i \eps} \right).
\label{eq:main-inel}
\end{align}
$\mathcal{P}_{n,2}$ is the set of all possibilities of dividing $n$ photons between the particles $(m_1,Q_1)$ and $(m_2,Q_2)$. 
For $r_i \neq 0$, the expression in (\ref{eq:main-inel}) is always $0$. $p_1, p_2$ and $r_i$ are on-shell. Since we require $p \cdot r = p_0 r_0 - \vec{p} \cdot \vec{r} = 0$, this implies $\sqrt{\vec{p}^{\:2} + m^2 } \, |\vec{r} \, | = |\vec{p} \, ||\vec{r} \, | \cos \tht$. This is only possible if $\cos \tht > 1$. Therefore, $\hat{\delta}(p.\ovr)$ is always zero for all on-shell values of $p$ and $r \neq 0$. Thus, the leading superclassical terms vanish. 
Any sub-leading corrections to the proper subset of massive propagators must also be $0$, which may be argued using the soft-photon theorem. Since the photon momentum scales with $\hbar$, the external photon momentum is soft in the classical limit. Thus, to find the $\hbar$ scaling of tree-level diagrams in $3+1$ dimensions, we can invoke the higher order soft-photon theorem proved in \cite{c:Inf-set-sof-thm-gg-th-wti} for complex scalar and photon interactions:
\begin{align} 
&\textit{A}(p_1, p_2 \to p_1', p_2', \Upsilon_n^{\vec{r},\vec{\sig}}) = \Bigg( \frac{1}{\omega^n}(S^{(0)})^n + \frac{1}{\omega^{n-1}} \left[ (S^{(0)})^{(n-1)}S^{(1)} + \text{contact} \right] 
\nonumber\\
& + \frac{1}{\omega^{n-2}} \left[ (S^{(0)})^{(n-1)}S^{(2)} + (S^{(0)})^{(n-2)}(S^{(1)})^2 + \text{contact} \right] +\ldots \Bigg) \textit{A}(p_1, p_2 \to p_1', p_2') + \text{Remainder},
\label{eq:h-soft-lim} 
\end{align} 
where
\begin{align} 
S^{(l)}\omega^{l-1} = \sum_{i=1}^{n_{ext}} (\eps_p)_\mu (r) \,
\frac{J_i^{\mu \nu}r_\nu}{p_i \cdot r} (r\cdot\dodo{}{p_i})^{l-1},
\label{eq:h-sof-fac}
\end{align} 
for $l \geq 1$. Here, $i$ denotes the external scalar leg that emits a photon with momentum $r$, and $J_i^{\mu \nu}$ denotes the angular momentum operator of the $i^{th}$ external scalar. $n$ represents the number of external photons and $n_{ext}$ is the number of external scalars. For the case considered here, $n_{ext}=4$. Contact terms contribute to the sub-leading soft factors and only appear when at least two external photons are present. They cannot be written solely as products of individual $S^{(i)}$ factors. They precisely include the terms containing $4$-point vertices, which we previously established to be sub-leading. To recall, a $4$-point vertex effectively replaces two $3$-vertices and an intermediate massive propagator. Since a massive propagator scales with an extra power of $1/\hbar$, its absence in a $4$-point vertex suppresses the factor, rendering it a sub-leading contribution. The remainder terms, on the other hand, correspond to the non-factorizable part of the amplitude. For example, in the case of a single external photon, the soft limit of such an amplitude can be decomposed into a sum of factorizable and non-factorizable terms as follows:
\begin{align}
\textit{A}(p_1, p_2 &\to p_1', p_2', \Upsilon_1^{\vec{r},\vec{\sig}}) 
\nonumber\\
&= \Bigg( \frac{1}{\omega}S^{(0)} + S^{(1)} 
+ \omega S^{(2)} +\ldots \Bigg) \textit{A}(p_1, p_2 \to p_1', p_2') +\underbrace{\omega R(p_1, p_2, p_1', p_2', \hat{r}_1, \sig_1)}_{\text{Remainder}}.
\end{align}
As shown in \cite{c:Inf-set-sof-thm-gg-th-wti}, the remainder terms appear in the sub-sub-leading order. For $n$ external photons, this implies that the remainder terms make their first appearance at $\mathcal{O}(1/\omega^{n-2})$.

For $l = 0$, Eqn.(\ref{eq:h-sof-fac}) simplifies to the leading soft-photon theorem \cite{c:wein-sof-th}:
\begin{align}
\frac{(S^{(0)})^n}{\omega^n} = \prod_{j=1}^{n} \left(\sum_{i=1}^{n_{ext}} i e Q_i
\frac{\eps_p (r_j) \cdot p_i}{p_i \cdot r_j} \right).
\label{eq:S0}
\end{align} 

Upon evaluating it for $n_{ext} = 4$, we obtain:
\begin{align}
\frac{(S^{(0)})^n}{\omega^n} = & \prod_{j=1}^{n} ie \left( 
\frac{Q_1 \, \eps_p(r_j) \cdot p_1}{p_1 \cdot r_j - i \eps} - \frac{Q_1 \, \eps_p(r_j) \cdot p_1}{p_1 \cdot r_j + i \eps} +  
\frac{Q_2 \, \eps_p(r_j) \cdot p_2}{p_2 \cdot r_j - i \eps} - \frac{Q_2 \, \eps_p(r_j) \cdot p_2}{p_2 \cdot r_j + i \eps_p}\right)
\nonumber\\
= & \prod_{j=1}^{n} \frac{(ie)^n}{\hbar^n} ( Q_1 \, \eps_p(r_j) \cdot p_1 \, \hat{\del}(p_1 \cdot \ovr_j) +  Q_2 \, \eps_p(r_j) \cdot p_2 \, \hat{\del}(p_2 \cdot \ovr_j)).
\label{eq:S00}
\end{align}
Using the KMOC formalism, if we scale the coupling constant $e$ with $1/\sqrt{\hbar}$ in Eqn.(\ref{eq:S00}), and if we take the classical limit of $A(p_1, p_2 \to p_1', p_2')$, we find that the first term in Eqn.(\ref{eq:h-soft-lim}) is the same as the expression in (\ref{eq:main-inel}). As argued before, for $r \neq 0$, these terms do not contribute. So, $S^{(0)}$ simply vanishes in Eqn.(\ref{eq:h-soft-lim}). Therefore, what we are left with are the contact terms and the remainder terms, which are shown to scale as the same power as the factorizable terms in \cite{grav-from-clas}. So, factorizable terms determine the $\hbar$ scaling. Since $S^{(0)}$ vanishes, all coefficients of $1/\omega^k$, for $n \geq k \geq 1$, vanish.

The first non-zero term appears when $\hbar$ corrections of
all $n$ massive propagators are considered. Thus, the amplitude at leading order must scale as $1/\hbar^{(n/2)+3}$. This implies $N( p_1, p_2 \to p_1', p_2', \Upsilon_n^{\vec{r},\vec{\sig}}) \sim 1/\hbar^{(n/2)+2}$ at $\hbar \to 0$. As we shall show, this contributes classically. Since the $1/\hbar^2$-scaling results from the exchange photon propagator, the extra $1/\hbar^{(n/2)}$ must stem from inserting n external photons. Thus, we may deduce that every insertion of an external photon in $N$ leads to an additional factor of $1/\sqrt{\hbar}$ . 
Using this in Eqn.(\ref{eq:chan}), assume that there are $k$ photon insertions between $N$ and $A_{n-1}^O$, then the extra factors due to photon insertions must scale as follows:
\begin{align} 
\underbrace{\hbar^{2k}}_{\text{Phase space measure}} \times \underbrace{\frac{1}{\hbar}}_{\text{From Eqn.(\ref{eq:recur})}} \times \underbrace{\frac{1}{\hbar^{k/2}}}_{\text{$k$ photon insertions in $N$}} \times \underbrace{\frac{1}{\hbar^{k/2}}}_{\text{$k$ photon insertions in $A_{n-1}^O$}} \sim \quad \hbar^{k-1}.
\label{eq:h-fac-inel}
\end{align}
Therefore, the classical contribution to $A_n^O$ due to inelastic channels can only come from $k=1$. For $k \geq 2$, $A_n^O$ is entirely quantum. Since the proof did not assume anything about the external states of $A_n^O$ in Eqn.(\ref{eq:chan}), Eqn.(\ref{eq:h-fac-inel}) generally holds, and $1/\hbar$ gets canceled for any matrix element of $A_n^O$ in the classical limit.

\subsection{Static Mode Contributions}
\label{0-freq-phot}
We have shown that the superclassical contributions from inelastic channels, consisting of photons with momenta $r \neq 0$, vanish. However, for $r=0$, the expression in (\ref{eq:main-inel}) can give superclassical contributions that violates Eqn.(\ref{eq:3pt1pf}). To see this, consider the factor of delta function in expression (\ref{eq:main-inel}):
\begin{align}
\hat{\del}(p \cdot r) = \hat{\del}(p_0 \, r_0 - \vec{p} \cdot \vec{r} \,) = \hat{\del}(|\vec{r} \, |(p_0 - \vec{p}\cdot\hat{r} )).
\label{eq:stat-modss}
\end{align}
$p_0$ cannot equal $\vec{p}\cdot\hat{r}$ for $|\vec{r} \, | \neq 0$. But $|\vec{r} \, | = 0$ can be a valid solution, leading to a non-zero contribution to the superclassical terms, potentially violating Eqn.(\ref{eq:3pt1pf}). This can be seen by rewriting the integration of Eqn.(\ref{eq:stat-modss}) over the phase space of static mode photons in polar coordinates as follows:
\begin{align}
 \hat{d}^{\,3} r \, \hat{\del}(|\vec{r} \, |(p_0- \vec{p}\cdot \hat{r} )) =   |\vec{r} \, |^2 \hat{d} |\vec{r} \, |\sin \tht \, \hat{d} \tht \hat{d} \phi \, \frac{\hat{\del}(|\vec{r} \, |)}{(p_0 - \vec{p}\cdot \hat{r} )}.
\label{eq:msure}
\end{align} 
In this section, we prove that static modes do not contribute and thus do not lead to any superclassicality. In other words, we prove that Eqn.(\ref{eq:3pt1pf}) still holds even when inelastic static mode channels are considered.  
We can show this by calculating the tree-level diagrams of $A_2^O$ for an inelastic channel with a single-photon insertion and by considering only contributions of the static modes in its classical limit. $A_2^O$ is given by: 
 \begin{align}
A_2^O = \frac{1}{\hbar}[N, A_{1}^O] = \frac{1}{\hbar^2}[N,[N,O]\,]. 
\label{eq:A2NNO}
\end{align} 
We insert a single photon between $N$ and $A_1^O$:
\begin{align}
 \la p_1', p_2' | A_2^O | p_1, p_2 \ra = \frac{1}{\hbar}\int d \phi(\tl{p}_1, \tl{p}_2, & \Upsilon_1^{\vec{r},\vec{\sig}}) (\la p_1', p_2' | N| \tl{p}_1, \tl{p}_2,\Upsilon_1^{\vec{r},\vec{\sig}}\ra \, \la \tl{p}_1, \tl{p}_2,\Upsilon_1^{\vec{r},\vec{\sig}} | A_1^O| p_1, p_2 \ra  
\nonumber\\
- & \la p_1', p_2' | A_1^O | \tl{p}_1, \tl{p}_2,\Upsilon_1^{\vec{r},\vec{\sig}} \ra \, \la \tl{p}_1, \tl{p}_2,\Upsilon_1^{\vec{r},\vec{\sig}} | N | p_1, p_2 \ra ),
\label{eq:locNpp}
\end{align}
where $p_1' = p_1+q$, $p_2' = p_2 - q$, $\tl{p}_1 = p_1 + q_+$ and $\tl{p}_2 = p_2 -q_+ - r$. Using Eqn.(\ref{eq:A1phot}) for $k=1$, we write $A_1^O$ in terms of $N$. We substitute the classical limit of the tree-level amplitude for $n=1$ in (\ref{eq:main-inel}) into $N(p_1, p_2 \to p_1', p_2', r, \sig )$ in Eqn.(\ref{eq:locNpp}). Evaluating Eqn.(\ref{eq:locNpp}) in the classical limit, we find:
\begin{align}
A_2^O&( p_1, p_2 \to  \, p_1', p_2') = 
\nonumber\\
& -\frac{8 \, \bar{e}^6 \, (Q_1 \, Q_2 \, p_1 \cdot p_2)^2 }{\hbar^4} \sum_p \int \hat{d}^4 \ovq_+ \:|\vec{\ovr}| \: \hat{d} |\vec{\ovr}|\sin \tht \: \hat{d} \tht \: \hat{d} \phi \: \hat{\del}^{(+)}(2 p_1\cdot \ovq_+) \,\hat{\del}^{(+)}(2 p_2 \cdot (\ovr+\ovq_+))
\nonumber\\
& \times \left(\frac{Q_1 \, p_1 \cdot \eps^*_p(r)}{(\ovq_++\ovr - \ovq)^2-i \eps} \frac{\hat{\del}(|\vec{\ovr}|)}{(p_1^0- \vec{p}_1 \cdot \hat{r} )} +  \frac{Q_2 \, p_2 \cdot \eps^*_p(r)}{(\ovq_+ - \ovq)^2-i \eps} \frac{\hat{\del}(|\vec{\ovr}|)}{(p_2^0- \vec{p}_2 \cdot \hat{r} )} \right) 
\nonumber\\
& \times \left(\frac{Q_1 \, p_1 \cdot \eps_p(r)}{(\ovq_++\ovr)^2+i \eps} \frac{\hat{\del}(|\vec{\ovr}|)}{(p_1^0- \vec{p}_1 \cdot \hat{r} )} +  \frac{Q_2 \, p_2 \cdot \eps_p(r)}{\ovq_+^2+i \eps} \frac{\hat{\del}(|\vec{\ovr}|)}{(p_2^0- \vec{p}_2 \cdot \hat{r} )} \right)
 \nonumber\\
 & \times \left( (2 \ovq_+^\mu  - \ovq^\mu) \nabla_\mu -  2 \ovr^\mu  \dodo{}{p_2^\mu} \right) O(p_1,p_2).
 \label{eq:A20-1ph}
\end{align}
The details of the calculation are given in the Appendix \ref{A:A2O-1ph}, where it is shown that, unlike the elastic channel, the leading term of $N A_1^O$  does not cancel the leading term in $A_1^O N$ for the inelastic channel in Eqn.(\ref{eq:locNpp}). In fact, they add up to give a common $N^2$ factor in the classical limit. Therefore, we have to compute Eqn.(\ref{eq:A20-1ph}) to determine whether the superclassical terms survive.
Integrating over $|\vec{r} \, |$ in Eqn.(\ref{eq:A20-1ph}), $\hat{\del}(|\vec{\ovr}|)$ in the integrand implies $|\vec{r}\, | = 0$. Since there is a product of two $\hat{\del}(|\vec{\ovr}|)$s, one from each $N$, this implies that $A_2^O$ for the leading order in $\hbar$ goes as $\sim  |\vec{\ovr}| \, \hat{\del}(|\vec{\ovr}|)$, where $|\vec{\ovr}| = 0$. So, $A_2^O$ is $0$ for the leading order. For sub-leading order $\sim 1/\hbar^3$ which is still superclassical, the tree-level amplitude contains a quantum correction due to a propagator or a vertex. Since any quantum correction to the vertex must be of the form: 
\begin{align}
2p^\mu + \hbar \al^\mu, 
\label{eq:vert-corr}
\end{align}
where $\al^\mu$ is the sum of the exchange momenta and the radiated momenta, $A_2^O$ with any quantum correction to the vertex must go as $|\vec{\ovr}|^s \, \hat{\del}(|\vec{\ovr}|)$, where $|\vec{\ovr}| = 0$ and $s$ is always an integer $\geq 1$. This ensures that the sub-leading superclassical terms in $A_2^O$ vanish for any quantum correction to the vertex. For quantum corrections to the propagator, we expand the propagator semi-classically:
\begin{align}
\frac{1}{(p+\hbar \al)^2 - m^2} = \frac{1}{2 \hbar p \cdot \al + \hbar^2 \al^2} = \frac{1}{2 \hbar p \cdot \al} -\frac{1}{4}\frac{\al^2}{(p \cdot \al)^2}.
\label{eq:corr-prop}
\end{align}
The first term in the expansion, $1/(2\hbar p \cdot \al + i\eps)$, is responsible for generating the delta function in expression (\ref{eq:main-inel}), defining the static modes. Since $\al^\mu$ is the sum of the exchange momenta and the radiated momenta, $A_2^O$ must go as $|\vec{\ovr}|^s$, where $s \geq 1 $. Thus, the sub-leading superclassical terms in $A_2^O$ vanish at $|\vec{\ovr}| = 0$. Here, the quantum correction to the massive propagator is considered only for one of the tree-level $N$ in Eqn.(\ref{eq:A2NNO}). For a more explicit representation, see Eqn.(\ref{eq:A2}). If both $N$ admit quantum corrections to the propagators in their tree-level diagrams, the corresponding term in $A_2^O$ will not contain any static modes, thereby reducing precisely to the classical limit discussed in the previous subsection. Thus, the contributions from the static modes always vanish and do not allow superclassical terms in $A_2^O$.

The conclusion can be generalized using Eqn.(\ref{eq:h-soft-lim}). For calculating $A_k^O$ in Eqn.(\ref{eq:chan}), assume that there is an insertion of $n$ photons. Using the expression in (\ref{eq:main-inel}), most superclassical static mode contributions can be shown to contain the following factors:
\begin{align}
 \int \, \prod_{l=1}^{n} \, \frac{\hat{d}^3 \ovr_l}{2 |\vec{\ovr_l}|}  \left(\prod_{l = 1}^{j_1}\hat{\del}(p_1 \cdot \ovr_l)\prod_{l = j_1+1}^{n}\hat{\del}(p_2 \cdot \ovr_l)\right) \left(\prod_{l = 1}^{j_2}\hat{\del}(p_1 \cdot \ovr_{l})\prod_{l = j_2+1}^{n}\hat{\del}(p_2 \cdot \ovr_l)\right)\ldots.
\end{align}
Using Eqn.(\ref{eq:msure}), this can be rewritten as:
\begin{align}
 \frac{1}{2^n}\int \, \prod_{l=1}^{n} |\vec{\ovr}_l| \, \hat{d} |\vec{\ovr}_l| \sin \tht_l \, \hat{d} \tht_l \, \hat{d} \phi_l  & \left(\prod_{l = 1}^{j_1}\frac{\hat{\del}(|\vec{\ovr}_l|)}{(p_1^0- \vec{p}_1 \cdot \hat{r} )}\prod_{l = j_1+1}^{n}\frac{\hat{\del}(|\vec{\ovr}_l|)}{(p_2^0- \vec{p}_2 \cdot \hat{r} )}\right) 
\nonumber\\
\times & \left(\prod_{l = 1}^{j_2}\frac{\hat{\del}(|\vec{\ovr}_l|)}{(p_1^0- \vec{p}_1 \cdot \hat{r} )}\prod_{l = j_2+1}^{n}\frac{\hat{\del}(|\vec{\ovr}_l|)}{(p_2^0- \vec{p}_2 \cdot \hat{r} )}\right) \ldots.
\label{eq:pprod-stat-mod}
\end{align}
   Integrating over $|\vec{\ovr}_l|$, we get $\prod_{l = 1}^{n} |\vec{\ovr}_l| \hat{\del}(|\vec{\ovr}_l|)$, which is $0$ at $|\vec{\ovr}_l| = 0$. Hence, static modes do not contribute to the leading superclassical order. For sub-leading superclassical orders in $\hbar$, corrections occur due to factors given in Eqn.(\ref{eq:h-sof-fac}). Any sub-leading superclassical term must replace one of the $\sim \hat{\del}(|\vec{\ovr}_l|) \to \sim |\vec{\ovr}_l|^s$, where $s \geq 0$. If all of them are replaced, then that will no longer be a static mode contribution. So, there must be at least one factor of $\hat{\del}(|\vec{\ovr}_l|)$ in Eqn.(\ref{eq:pprod-stat-mod}). This must lead to a factor of $\int|\vec{\ovr}_l|^{s+1} \, \hat{d} |\vec{\ovr}_l| \, \hat{\del}(|\vec{\ovr}_l|)$, which is $0$ since $s \geq 0$. Therefore, static modes do not contribute. Since we had shown in the last section \S\ref{s:in-cc} that any superclassical terms for inelastic channels could only result due to static modes, the lack of contributions of static modes shown in this section establishes $A_k^O$ is superclassical free.
   
\section{Classical Limit of Angular Impulse} 

In this section, we analyse the classical angular impulse using the KMOC formalism. Our goal, as before, is to prove that in exponential representation, the superclassical terms vanish to all orders in perturbation theory. 
The orbital angular momentum operator for a scalar particle with momentum $p^{\mu}$ is defined as:
\begin{align}
O = L_{\mu \nu} =i \hbar\, (p_\mu \partial_{\nu} - p_\nu \partial_{\mu} ) = i \hbar\, ( p \wedge \dodo{}{p} )_{\mu\nu}.
\end{align}
The angular impulse for the particle-$1$ is then obtained by the following formula:
\begin{align}
\la \Delta L_{1} \ra (p_1, p_2, b) = \lim_{\hbar \to 0}  \int \hat{d}^4 p_1' \hat{d}^4 p_2' \: \hat{\del}(p_1'^2 - m_1^2) \hat{\del}(p_2'^2 - m_2^2)  \: e^{-ib \cdot \ovq} \, \la p_1', p_2' |\Delta L_1| p_1, p_2 \ra,
\label{eq:clas-exp-2}
\end{align}
where using Eqn.(\ref{eq:D-O-N_opt}), $\Delta L_1$  is given by:
\begin{align}
\Delta L_1 = \sum_{n \geq 1} \frac{(-i)^n}{\hbar^n n!} [N,[N,\ldots[N,L_1]]] = \sum_{n \geq 1} \frac{(-i)^n}{ n!}A_n^{L_1}.
\label{eq:D-L-N_opt}
\end{align}
We now prove the following:
\\
\framedtext{
\begin{align}
 \la \Delta L_{1} \ra (p_1, p_2, b) \sim \hbar^0. 
 \end{align}
 }
 \\
 \\
As with the linear impulse, the proof for the angular impulse is divided into two parts. First, we provide the proof for the elastic channel contributions in the intermediate channel. The proof for the inelastic channel contributions follows the same arguments as those for the linear impulse. To begin, we evaluate $A_{1}^{L_{1}}$ for the elastic channel:
\begin{align}
 \la p_1', p_2' | A_1^{L_1} | p_1, p_2\ra
& = \frac{1}{\hbar}\la p_1', p_2' |[N, i \, \hbar \, \hat{p_1} \wedge \hat{\dodo{}{p_1}}]| p_1, p_2 \ra 
\nonumber\\
& =  \, i\, p_1 \wedge \dodo{}{p_1} ( N(p_1 \cdot p_2, (p_1' - p_1)^2) \, \hat{\delta}^{(4)}(p_1'+ p_2' - p_1 - p_2)) 
\nonumber\\
& + i \,  p_1' \wedge \dodo{}{p_1'}( N(p_1 \cdot p_2, (p_1'-p_1)^2) \, \hat{\delta}^{(4)}(p_1'+ p_2' - p_1 - p_2)).
\label{eq:A1L1}
\end{align}
We use the Lorentz covariant representation of $N$, defined in Appendix \ref{ap:el-ch-cny}, for the elastic channel.
Using the Lorentz covariance of the angular momentum operator and energy-momentum conservation, we can simplify the above identity as follows:
\begin{align}
\la p_1', p_2' |\Delta L_{1}| p_1, p_2 \ra  & =  \, ( N( p_1' \cdot p_2', (p_2' - p_2)^2) p_1 \wedge \dodo{}{p_1} 
\nonumber\\
 & + N(p_1 \cdot p_2, (p_2' - p_2)^2) \, p_1' \wedge \dodo{}{p_1'} ) \, \hat{\delta}^{(4)}(p_1'+ p_2' - p_1 - p_2). 
\label{eq:delLL}
\end{align}
Writing $p_1' = p_1 + q_1$ and $p_2' = p_2 + q_2$, and evaluating $N$ in the semi-classical expansion, we find:
\begin{align}
 N(p_1 \cdot p_2  +  \hbar \, p_1 \cdot \ovq_2 + \hbar \, p_2 \cdot \ovq_1 + & \hbar^2 \, \ovq_1 \cdot \ovq_2, \, \hbar^2 \, \ovq_2^2) = (1 + \hbar \, \Box\,) \, N(p_1 \cdot p_2, \, \hbar^2 \, \ovq_2^2),
 \nonumber\\
 & \text{where} \, \, \, \Box = (\ovq_2\cdot \dodo{}{p_2} + \ovq_1 \cdot \dodo{}{p_1}). 
\label{nexpinhbar}
\end{align}
Substituting Eqn.(\ref{nexpinhbar}) into Eqn.(\ref{eq:delLL}), we find that the superclassical term cancels and the classical contribution to $A_{1}^{L_{1}}$ is given by:
\begin{align}
 \la p_1', p_2' |\Delta L_1| p_1, p_2 \ra = \, &\hbar\,\left( p_1 \,\Box -  \ovq_1\right) N(p_1 \cdot p_2, \, \hbar \ovq_2^2) \wedge \dodo{}{p_1} \hat{\delta}^{(4)}(p_1'+ p_2' - p_1 - p_2).
 \label{eqn:4pt8}
\end{align}
We can finally substitute $A_{1}^{L_{1}}$ within the KMOC formula, Eqn.(\ref{eq:clas-exp-2}). We find:
\begin{align}
\la \triangle L_{1} \ra (p_1, p_2, b) = \lim_{\hbar \to 0} \frac{\hbar^2}{2} \int \hat{d}^4 \ovq_1 \:\hat{d}^4 \ovq_2 \: \hat{\del}^4(\ovq_1 + \ovq_2)\:( \,p_1 \wedge \triangle_{\ovq} \, (g \, \Box \, N) - \ovq_1 \triangle_{\ovq} (g N)\,),
\end{align}
where
\begin{align}
 \bigtriangleup_{\ovq} = \dodo{}{\ovq_1}+\dodo{}{\ovq_2}, \, \,\text{and} \, \, g = \hat{\del}(2 p_1 \cdot \ovq_1) \, \hat{\del}(2 p_2 \cdot \ovq_2)  \: e^{- ib \cdot \ovq_1}. 
\end{align}
$N$ is a function $ N(p_1 \cdot p_2, \, \hbar^2 \ovq_2^2)$.

As $N(p_1 \cdot p_2, \, \hbar^2 \ovq_2^2) \sim 1/\hbar^2$, it follows that $\triangle L_{1}\, \sim\, \hbar^0$. Thus, we see that the contribution from $A_1^{L_1}$ is smooth in the classical limit. 

We extend this result to all orders in $\hat{N}$. As before, our proof proceeds via induction. 
Assume that $A _{n-1}^{L_{1}}$ has a smooth classical limit. As shown in Appendix \ref{a:generic-anl1}, $A_{n-1}^{L_{1}}$ has the following form:
\begin{align}
\la p_1', p_2' | A_{n-1}^{L_1} | p_1, p_2 \ra =  (\,A_{n-1}^{L_1} (p_1, p_2, q_1)+ A_{n-1}^{L_1} (p_1, p_2, q_1, q_2) \wedge \dodo{}{p_1}\,)\,\hat{\delta}^{(4)}(p_1'+ p_2' - p_1 - p_2),
\label{eq:ang.ansz}
\end{align}
where $p_1' = p_1 + q_1$ and $p_2' = p_2 + q_2$. The cancellation of superclassical terms for the first term follows the same procedure as for the linear impulse. So, we focus on showing the cancellation of superclassical terms for the second term. Using the recursion for $A_{n}^{L_1}$, we have: 
\begin{align}
\la p_1', p_2' | A_n^{L_1} | p_1, p_2 \ra = \frac{1}{\hbar} \la p_1', p_2' | [N, A_{n-1}^{L_1}] | p_1, p_2 \ra.
\label{eq:AnLmunu}
\end{align}
We evaluate the above quantity in the elastic channel. We divide the commutator into two terms:
\begin{align}
f_{L_{1}} =\frac{1}{\hbar} \la p_1', p_2' | N A_{n-1}^{L_1}| p_1, p_2 \ra, \quad \text{and} \quad
f^{\prime}_{L_{1}} = \frac{1}{\hbar} \la p_1', p_2' | A_{n-1}^{L_1} N | p_1, p_2 \ra.
\end{align}
Evaluating $f_{L_1}$, by inserting elastic channel, $\int  d \phi(\tl{p}_1) d \phi(\tl{p}_2) | \tl{p}_1,\tl{p}_2 \rangle \langle \tl{p}_1,\tl{p}_2 |$, between $N$ and $A_{n-1}^{L_1}$, we obtain:
\begin{align}
f_{L_1} = \frac{1}{\hbar} \int &  d \phi(\tl{p}_1) d \phi(\tl{p}_2) \, N (\tl{p}_1.\tl{p}_2, (p_1'-\tl{p}_1)^2) \, \hat{\delta}^{(4)}(p_1'+p_2' - \tl{p}_1- \tl{p}_2) 
\nonumber\\ 
 & \times A_{n-1}^{L_1}(p_1, \, p_2, \, \tl{p_1}- p_1, \tl{p_2}- p_2) \wedge \dodo{}{p_1}\hat{\delta}^{(4)}(\tl{p}_1+\tl{p}_2 - p_1 - p_2).
\end{align}
Let $\tl{p}_1 = p_1 + q_+$ and $\tl{p}_2 = p_2 + q_-$. We use the on-shell condition and 4-momentum conservation to get $\tl{p}_1 \cdot \tl{p}_2 = p_1' \cdot p_2'$. We change the variables to $ 2 q' = q_+ + q_- $ and $2 q = q_+ - q_- $. Then, we integrate over $q'$ to obtain:
\begin{align}
-\frac{1}{2 \hbar}\int \hat{d}^4 q & \: \hat{\del}((p_1+\frac{q_1 + q_2}{2}+ q)^2 - m_1^2) \, \hat{\del}((p_2+\frac{q_1 + q_2}{2} - q)^2 - m_2^2) 
\nonumber\\
& \times N \left(p_1' \cdot p_2', \left(q_2+q-\frac{q_1 + q_2}{2}\right)^2 \right)
\nonumber\\
&\times A_{n-1}^{L_1}\left(p_1, \, p_2, \, q + \frac{q_1 + q_2}{2}, - q + \frac{q_1 + q_2}{2}\right) \wedge \dodo{}{q'}\hat{\delta}^{(4)}(2q')\bigg|_{q' =(q_1+q_2)/2}.
\label{eq:comstp}
\end{align}

Changing $q \to - q + (q_1 - q_2)/2$, we find:
\begin{align}
f_{L_{1}} = -\frac{1}{2\hbar}\int \hat{d}^4 q & \: \hat{\del}((p_1 - q + q_1)^2 - m_1^2) \, \hat{\del}((p_2+ q + q_2)^2 - m_2^2) \, N (p_1' \cdot p_2', q^2)
\nonumber\\
&\times A_{n-1}^{L_1}(p_1, \, p_2, \, q_1-q, \, q_2 + q) \wedge \dodo{}{q'}\hat{\delta}^{(4)}(2q')\bigg|_{q' =(q_1+q_2)/2}.
\label{eq:ttl-L-1'}
\end{align} 
We similarly compute $f^{\prime}_{L_{1}}$ employing the same steps to obtain Eqn.(\ref{eq:comstp}). Upon doing so, we get the following:
\begin{align}
 f^{\prime}_{L_{1}} = \frac{1}{2\hbar} \int & \hat{d}^4 q  \: \hat{\del}((p_1+ q)^2 - m_1^2) \, \hat{\del}((p_2 - q)^2 - m_2^2) \, N (p_1 \cdot p_2, q^2)  
\nonumber\\
&\times A^{L_1}_{n-1}(p_1+q, p_2 - q, q_1 - q, q_2 + q) \wedge \dodo{}{q'}\hat{\delta}^{(4)}(q_1+q_2 - 2 q') \bigg|_{q'=0}.
\label{eq:ttl-L-2'}
\end{align}
At $\hbar \to 0$, both the above expressions in Eqn.(\ref{eq:ttl-L-1'}) and Eqn.(\ref{eq:ttl-L-2'}) reduce to:
\begin{align}
f_{L_{1}} = f^{\prime}_{L_{1}} = \hbar \int & \hat{d}^4 \ovq \: \hat{\del}(2p_1 \cdot \ovq +  \hbar \, \ovq^2) \, \hat{\del}(2p_2 \cdot \ovq -  \hbar \ovq^2) \, N (p_1 \cdot p_2, \hbar^2 \,\ovq^2) 
\nonumber\\
& \times A_{n-1}^{L_1}(p_1, p_2, \hbar(\ovq_1- \ovq), \hbar(\ovq_2 + \ovq)) \wedge \dodo{}{p_1}\hat{\delta}^{(4)}(p_1'+ p_2' - p_1 - p_2).
\label{eq:ttl-L-3}
\end{align}
Thus, the leading order in $f_{L_{1}}$ cancels with the leading order in $f^{\prime}_{L_{1}}$, ensuring the cancellation of superclassical terms. The extra factor of $1/\hbar$ in Eqn.(\ref{eq:AnLmunu}) thus cancels. Note that we have retained an $\hbar$ correction within the delta function in Eqn.(\ref{eq:ttl-L-3}) in order to isolate those terms that cancel at the sub-leading order, leaving only non-zero contributions to the classical limit. To obtain the classical term, we expand the terms in Eqn.(\ref{eq:ttl-L-1'}) and Eqn.(\ref{eq:ttl-L-2'}). The delta functions in Eqn.(\ref{eq:ttl-L-1'}) can be expanded as follows:
\begin{align}
\hat{\del}(2p_1 \cdot \ovq + \hbar  \ovq^2 + 2 \hbar  \ovq.(\ovq_1 -  \ovq)) = 
(1 + \hbar \,(\Box -  \ovq^\mu \nabla_{\mu}))\hat{\del}(2p_1 \cdot \ovq +  \hbar  \ovq^2),
\nonumber\\ 
\hat{\del}(2p_2. \ovq - \hbar  \ovq^2 + 2 \hbar  \ovq.(\ovq_2 +  \ovq)) =  (1 + \hbar \,(\Box -  \ovq^\mu \nabla_{\mu}))\hat{\del}(2p_2 \cdot \ovq -  \hbar \ovq^2).
 \end{align}
The $N$ in Eqn.(\ref{eq:ttl-L-1'}) admits the same expansion as in Eqn.(\ref{nexpinhbar}). Similarly, $A^{L_1}_{n-1}$ in Eqn.(\ref{eq:ttl-L-2'}) can be expanded as:
\begin{align}
 A^{L_1}_{n-1}(p_1+\hbar \ovq, \, p_2 - \hbar \ovq, \, & \hbar(\ovq_1 - \ovq), \,  \hbar (\ovq_2 + \ovq)) =
 \nonumber\\
  & (\,1 + \hbar \ovq^\mu \nabla_{\mu}) \, A^{L_1}_{n-1}(p_1, \, p_2, \, \hbar(\ovq_1 - \ovq), \, \hbar (\ovq_2 + \ovq)).
\label{eq:expns-22}
\end{align}
Substituting them into the second term of Eqn.(\ref{eq:ang.ansz}) for $A_n^{L_1}$, we obtain:
\begin{align}
 A_n^{L_1} (p_1, p_2, q_1,q_2) = \nonumber \\
  \hbar^2 \int \hat{d}^4 \ovq \:  \Box  \, (\hat{\del}(2p_1 \cdot \ovq )& \, \hat{\del}(2p_2\cdot \ovq ) \, N (p_1 \cdot p_2, \, \hbar^2 \ovq^2))
 A_{n-1}^{L_1}(p_1, \, p_2, \, \hbar(\ovq_1- \ovq), \, \hbar(\ovq_2 + \ovq)) \nonumber \\
 - \hbar^2 \int \hat{d}^4 \: \ovq \,  \ovq^\mu \nabla_\mu  \, (\hat{\del}(2p_1 &\cdot \ovq) \, \hat{\del}(2p_2 \cdot \ovq) \, A_{n-1}^{L_1}(p_1, \, p_2, \, \hbar(\ovq_1- \ovq), \, \hbar(\ovq_2 + \ovq)))
 N (p_1 \cdot p_2, \, \hbar^2 \ovq^2)  ,
\nonumber\\
\text{where} \quad
  \Box \, = \, (\, \ovq_2^\mu & \dodo{}{p_2^\mu} + \ovq_1^\mu\dodo{}{p_1^\mu} \,)
 \quad \text{and} \quad
\nabla_\mu = (\, \dodo{}{p_1^\mu} - \dodo{}{p_2^\mu} \, ).
\label{eq:ang.ansz.finr}
\end{align}
Since $N (p_1 \cdot p_2, \, \hbar^2 \ovq^2) \sim 1/\hbar^2$ and $A_{n-1}^{L_1}$ has a smooth classical limit, it follows from Eqn.(\ref{eq:ang.ansz.finr}) that $A_n^{L_1}$ also admits a smooth classical limit. 

Next, we consider the inelastic channel contributions in Eqn.(\ref{eq:AnLmunu}). For the inelastic matrix elements of $A_1^{L_1}$, the Lorentz covariance of the angular momentum operator and the energy-momentum conservation can again be used to simplify Eqn.(\ref{eq:A1L1}) to Eqn.(\ref{eq:delLL}). This expression can then be evaluated in the semi-classical expansion to show the cancellation of the superclassical terms, similar to the way we obtained Eqn.(\ref{eqn:4pt8}). For $n \geq 2$, the cancellation of the superclassical terms in $A_n^{L_1}$  follows from the same arguments as those used for the linear impulse in \S\ref{s:in-cc} and \S\ref{0-freq-phot}. All cancellations occur at the level of evaluation of the $N$ itself. The tree-level diagram determines the $\hbar$ scaling of $N$, whose superclassical terms only allow static modes. The static modes are shown to vanish in their contribution to the final observable. Thus, no superclassical terms survive.

\section{Classical Limit of Radiation}

In this section, we analyse the classical limit of the four-vector potential using the KMOC formalism and show that all superclassical terms vanish in the exponential representation.
The momentum-space four-vector potential is defined by:
\begin{align} 
O = A_{\mu} (k) = \sqrt{2 E_k} \sum_r a_k^r \eps^r_\mu (k) + a_k^{r\dagger} \eps^{r*}_\mu (k), 
\end{align}
where $k$ is on-shell with $k^0 > 0$, and $r$ denotes the polarization. We prove that the classical limit of $A_{\mu} (k)$ is free of superclassical terms, i.e.,
\framedtext{
\begin{align}
\lim_{\hbar \to 0} \hbar^{3/2} \la A_{\mu} (k) \ra (p_1, p_2, b) \sim \hbar^0,
\label{eq:amuk}
\end{align}
}
 where $\hbar^{3/2} \la A_{\mu} (k) \ra (p_1, p_2, b)$ is given by:
 \begin{align}
\hbar^{3/2} \la A_{\mu} (k) \ra (p_1, p_2, b) = \lim_{\hbar \to 0} \hbar^{3/2} \int \hat{d}^4 p_1' \hat{d}^4 p_2' \: \hat{\del}(p_1'^2 - m_1^2) \hat{\del}(p_2'^2 - m_2^2)  \: e^{-ib \cdot \ovq} \, \la p_1', p_2' | \Delta A_{\mu} (k)| p_1, p_2 \ra.
\label{eq:clas-exp-3-A-f}
\end{align}
Using Eqn.(\ref{eq:D-O-N_opt}), $\Delta A_{\mu} (k)$  is given by:
\begin{align}
\Delta A_{\mu} (k) = \sum_{n \geq 1} \frac{(-i)^n}{\hbar^n n!} [N,[N,\ldots[N,A_{\mu} (k)]]] = \sum_{n \geq 1} \frac{(-i)^n}{ n!}A_n^{A_{\mu} (k)}.
\label{eq:D-Amu-N_opt}
\end{align}
As we shall see, this is equivalent to proving:
\framedtext{
 \begin{align}
\hbar^{3/2} \, \la p_1', p_2' | A_n^{A_{\mu} (k)} | p_1, p_2 \ra \sim \frac{1}{\hbar^2} \hat{\del}(p_1'+p_2'-p_1-p_2 \pm k).
\label{eq:A_n^A_mu(k)}
\end{align} 
}
\\
\\
The factor of $1/\hbar^2$ is expected in Eqn.(\ref{eq:A_n^A_mu(k)}) for Eqn.(\ref{eq:amuk}) to be classical. This is because when the delta function in Eqn.(\ref{eq:A_n^A_mu(k)}) is integrated in Eqn.(\ref{eq:clas-exp-3-A-f}), it yields an extra factor of $\hbar^2$,  identical to the scaling in Eqn.(\ref{eq:clas-exp}).

We carry out the proof via induction.
First, we calculate $ \la p_1', p_2' | A_1^{A_{\mu}(k)} | p_1, p_2 \ra   $:
\begin{align}
 \la p_1', p_2' | A_1^{A_{\mu}} (k)| p_1, p_2 \ra & =  \frac{\sqrt{2 E_k}}{\hbar}  \sum_r \, \la p_1', p_2' | [N, a_k^r \eps^r_\mu (k) +  a_k^{r\dagger} \eps^{r*}_\mu (k)] | p_1, p_2 \ra 
\nonumber\\
& =  \frac{1}{\hbar}\sum_r - \, \la p_1', p_2', k,r | N  | p_1, p_2 \ra \: \eps^r_\mu (k) + \la p_1', p_2'| N  | p_1, p_2, k,r  \ra \: \eps^{r*}_\mu (k), 
\label{eq:A1Oa}
\end{align}
where 
\begin{align}
\la & p_1', p_2', k,r | N  | p_1, p_2 \ra =  \, N(p_1, p_2 \to  p_1', p_2', k,r) \, \hat{\delta}^{(4)} (p_1' + p_2' + k - p_1 - p_2), 
\nonumber\\
& \text{and}\,\,
\la p_1', p_2'| N  | p_1, p_2, k,r  \ra =  \, N(p_1, p_2, k,r \to p_1', p_2') \, \hat{\delta}^{(4)} (p_1' + p_2' - p_1 - p_2 - k).
\label{eq:fact-del}
\end{align}
For the tree level contribution to $N$, the leading order in $\hbar$ is given by: 
\begin{align}
N(p_1, p_2 \to p_1', p_2', k, r) = i \frac{8 \, \bar{e}^3 \, p_1 \cdot p_2}{\hbar^{7/2}} \left(\frac{p_1 \cdot \eps^r(k)}{(\bar{q}_1+\ovk)^2+i \eps} \, \hat{\del}(2 p_1 \cdot \ovk) +  \frac{p_2 \cdot \eps^r(k)}{\bar{q}_1^2+i \eps} \, \hat{\del}(2 p_2 \cdot \ovk)\right),
\label{eq:N2to3}
\end{align}
where $p_1' = p_1 + q_1$ and $p_2' = p_2 - q_1 - k$ for the first case in Eqn.(\ref{eq:fact-del}).
The above expression is $0$ for $\ovk \neq 0$. Therefore, the first non-zero term for $k \neq $0 appears at 
\begin{align}
N(p_1, p_2 \to p_1', p_2', k, r) \sim \frac{\bar{e}^3}{\hbar^{5/2}}.
\label{eq:1use}
\end{align}
Since $N$ is Hermitian, $N(p_1, p_2, k, r \to p_1', p_2') = N^*(p_1', p_2' \to p_1, p_2, k, r)$, which scales identically as given in Eqn.(\ref{eq:1use}). 
 Using this in Eqn.(\ref{eq:A1Oa}), we get:
\begin{align}
 A_1^{A_{\mu}(k)}( p_1, p_2 \to p_1', p_2' ) \sim \frac{1}{\hbar^{7/2}}.
\end{align}
 Thus, 
\begin{align}
 \hbar^{3/2}  A_1^{A_{\mu}(k)}( p_1, p_2 \to p_1', p_2' ) \sim \frac{1}{\hbar^2},
\end{align}
which establishes that $A_1^{A_{\mu}(k)}$ is free from superclassical terms in the limit $\hbar \to 0$. Next, we use induction to investigate whether this holds for $A_n^{A_{\mu}(k)}$. We assume $\hbar^{3/2} A_{n-1}^{A_{\mu}(k)}(p_1, p_2 \to p_1', p_2') \sim 1/\hbar^2$ and prove the same for $A_n^{A_{\mu}(k)}$. We begin by evaluating the elastic channel:
\begin{align}
\la p_1', p_2' | A_n^{A_{\mu}(k)} | p_1, p_2 \ra =  \frac{1}{\hbar} \int & \hat{d}^4 \tl{p}_1 \hat{d}^4\tl{p}_2 \, \hat{\del}(\tl{p}_1^2 - m_1^2) \hat{\del}(\tl{p}_2^2 - m_2^2) \, ( \la p_1', p_2' | N| \tl{p}_1, \tl{p}_2 \ra 
\nonumber\\
& \times \la \tl{p}_1, \tl{p}_2 | A_{n-1}^{A_{\mu}(k)}| p_1, p_2 \ra  
 - \la p_1', p_2' | A_{n-1}^{A_{\mu}(k)} | \tl{p}_1, \tl{p}_2 \ra \la \tl{p}_1, \tl{p}_2 | N | p_1, p_2 \ra).
\label{eq:elas-ch-rad}
 \end{align}
Here, too, we label the two terms of commutator as:
\begin{align}
f_{A_{\mu}(k)} =\frac{1}{\hbar} \la p_1', p_2' | N A_{n-1}^{A_{\mu}(k)}| p_1, p_2 \ra, \quad
f^{\prime}_{A_{\mu}(k)} = \frac{1}{\hbar} \la p_1', p_2' | A_{n-1}^{A_{\mu}(k)} N | p_1, p_2 \ra.
\end{align}
 Before proceeding further, we state the following identities. Assume $O$ is a general operator and not necessarily Hermitian. Then, the following holds:
\begin{align}
 & A_n^{O_1+O_2} =  \, A_n^{O_1} + A_n^{O_2} \quad \text{and} \quad (A_n^O)^{\dagger} = (-1)^n A_n^{O^\dagger}
\nonumber\\
& \implies  A_n^{O+O^\dagger} =  \, A_n^O + A_n^{O^\dagger} = A_n^O + (-1)^n(A_n^O)^{\dagger}.
\label{eq:herm}
\end{align}
Since $A_{\mu} (k) = O + O^\dagger = \sqrt{2 E_k} \sum_r a_k^r \eps^r_\mu (k) + a_k^{r\dagger} \eps^{r*}_\mu (k)$,
we can use Eqn.(\ref{eq:herm}) to write:
\begin{align}
\la p_1', p_2' | A_n^{O+O^\dagger} | p_1, p_2 \ra = \la p_1', p_2' | A_n^O | p_1, p_2 \ra + (-1)^n (\la p_1, p_2 | A_n^O | p_1', p_2' \ra)^*,
\label{eq:Aedel}
\end{align}
where $O = \sqrt{2 E_k} \sum_r a_k^r \eps^r_\mu (k)$.
Factoring the delta functions, we get the following:
\begin{align}
\la p_1', p_2' | A_n^{O+O^\dagger} | p_1, p_2 \ra = & \sum_r A_n^O(p_1, \, p_2, \, q_1 , \, k , \, r) \, \eps^r_\mu (k) \: \hat{\del}^{(4)}(p_1'+p_2'+k - p_1 - p_2) 
\nonumber\\
& + (-1)^n \, (A_n^O(p_1', \, p_2', \, -q_1, \, k , \, r))^* \, \eps^{r*}_\mu (k) \: \hat{\del}^{(4)}(p_1'+p_2' - p_1 - p_2 - k),
\label{eq:Aedel-1}
\end{align}
where $p_1' = p_1 + q_1$ and $p_2' = p_2 + q_2$. Showing the cancellation of superclassical terms in the first term of Eqn.(\ref{eq:Aedel-1}) suffices to establish a smooth classical limit for radiation. Therefore, we substitute the first term for $A_{n-1}$ in Eqn.(\ref{eq:Aedel-1}) into Eqn.(\ref{eq:elas-ch-rad}) to determine $A_n^O ( p_1 , p_2, \hbar  \ovq_1, \hbar \ovk,r )$. Upon substitution, we find: 
\begin{align}
 A_n^O ( p_1 , p_2, & \hbar  \ovq_1, \hbar \ovk,r )  = 
  -\hbar^2  \int \hat{d}^4 \ovq_+ (\hat{\del}(2p_1\cdot \ovq_+) \hat{\del}(2p_2 \cdot (\ovq_++\ovk)) \, A_{n-1}^O(p_1, p_2, \hbar \ovq_+, \hbar \ovk , r) 
\nonumber\\
 & \times  \ovk^\mu\bigtriangleup_\mu N(p_1 \cdot p_2, \hbar^2 (\ovq_1 - \ovq_+)^2)
 +(\ovq_1 - \ovq_+ )^\mu\nabla_\mu( \hat{\del}(2p_1\cdot \ovq_+) \hat{\del}(2p_2 \cdot (\ovq_++\ovk)) 
 \nonumber\\
 &\times A_{n-1}^O(p_1, p_2, \hbar \ovq_+, \hbar \ovk , r)) \, N(p_1 \cdot p_2, \hbar^2 (\ovq_1 - \ovq_+)^2)), 
 \nonumber\\
& \text{where} \quad \bigtriangleup_\mu = \dodo{}{p_1^\mu}+  \dodo{}{p_2^\mu} \quad \text{and} \quad \nabla_\mu = \dodo{}{p_1^\mu} - \dodo{}{p_2^\mu}.
\label{eq:class-rad}
\end{align}
The details of the computation are given in Appendix \ref{Ap:rad-elas}. Similarly to the previous calculation for the linear and angular momentum impulses, we find that the leading order terms, which are superclassical, cancel. The terms in the next order in $\hbar$ cancel the extra factor of $1/\hbar$ in Eqn.(\ref{eq:elas-ch-rad}), leading to Eqn.(\ref{eq:class-rad}), which contributes classically. To show this, note that:
\begin{align}
\lim_{\hbar \to 0} \hbar^{3/2} A_n^{A_{\mu}(k)} \sim \hbar^2 \hbar^{3/2} A_{n-1}^{A_{\mu}(k)}(p_1, p_2 \to p_1', p_2')  N(p_1, p_2 \to p_1', p_2') \sim \hbar^2 \hbar^{3/2} \frac{1}{\hbar^{7/2}}\frac{1}{\hbar^{2}} \sim \frac{1}{\hbar^2}.
\end{align} 
For the inelastic channel, we examine an arbitrary matrix element of $\hbar^{3/2}A_1^O$. We find $ \hbar^{3/2}\la n | A_1^O | l \ra \sim  \sqrt{\hbar}( \la n | N a_k^r  | l \ra - \la n | a_k^r N  | l \ra )$. In the first term,  $a_k^r$ annihilates a photon, whereas in the second term,  $a_k^r$ creates a photon. Thus, as $\hbar \to 0$,  $ - \sqrt{\hbar}\,\la n |  a_k^r N  | l \ra $ dominates. Now, as shown in \S\ref{s:in-cc}, any external photon is responsible for a factor of $1/\sqrt{\hbar}$, we find that no factor of $1/\hbar$ survives in $\hbar^{3/2} A_1^O$. In the case of $\hbar^{3/2} A_n^O$ for $n \geq 2$, the arguments in \S\ref{s:in-cc} and \S\ref{0-freq-phot} still hold.

The result also extends to: 
\begin{align}
\lim_{\hbar \to 0} \hbar^{3m/2} \la A_{\mu}(k_1) \, A_{\mu}(k_2)\ldots A_{\mu}(k_m)\ra (p_1, p_2, b). 
\end{align}
The above term inserts a maximum of $m$ photons. Since the creation of each photon introduces an overall factor of $1/\sqrt{\hbar}$ in the classical limit, we obtain the following:
\begin{align}
 \hbar^{3m/2}  A_1^{A_{\mu}^m (k)} ( p_1, p_2 \to p_1', p_2' ) \sim  \hbar^{3m/2} \frac{1}{\hbar} N(p_1, p_2 \to p_1', p_2',\Upsilon_m^{\vec{r},\vec{\sig}}) \sim \frac{1}{\hbar} \hbar^{3m/2}\frac{1}{\hbar^{m/2}} = \hbar^{m-1}.
\end{align}
Therefore, for $m \geq 2$, there is no classical contribution and the expression is purely quantum. The induction argument straightforwardly extends to $A_n^O$, as it contains terms such as:
\begin{align}
\la p_1', p_2' | A_n^{A_{\mu}^m (k)} | p_1, p_2 \ra \sim  \sum_r A_n^O(p_1, p_2, q, k_1 , r_1, k_2, r_2\ldots k_m , r_m)
\nonumber\\
 &\hspace*{-2in} \times \prod_i \eps^{r_i}_\mu (k) \: \hat{\del}^{(4)}(p_1'+p_2'- p_1 - p_2 + \sum_i k_i).
\label{eq:Aedel-2}
\end{align}
Hence, the same argument for Eqn.(\ref{eq:Aedel-1}) follows in this case with the total photon momenta $\sum_i k_i$, which preserves the order of $\hbar$ in $A_n^{A_{\mu}^m (k)}(p_1, p_2 \to p_1', p_2')$ relative to $A_1^{A_{\mu}^m (k)} ( p_1, p_2 \to p_1', p_2' )$, making it purely quantum for $m \geq 2$. The contribution of the inelastic channel also follows in a similar fashion, making it purely quantum for $m \geq 2$.

\section{Discussion}

In this paper, we have provided proof that the classical limit in the KMOC formalism is smooth to all orders in perturbation theory for the three classes of asymptotic observables, i.e., the scattering angle, the angular impulse, and the radiative electromagnetic field. For the scattering angle, all order proof was carried out in \cite{exp-S-Mat-KMOC} for the elastic channels. We extended the proof to include the inelastic channels, completing the proof for the smoothness of the classical limit for linear impulse to all orders. We further proved the same for the angular impulse and the radiative electromagnetic field. This problem is important to address since the KMOC formalism has not been derived from first principles, and there is no direct way to argue the well-defined behaviour of the classical results without actually computing the amplitudes. In fact, one finds potentially uncontrolled divergences growing at each order of coupling constant, making the classical limit itself unattainable, even if we allow freedom of overall scaling. So, for universal applicability of the KMOC formalism, it becomes necessary to show that the superclassical terms are absent to all orders in perturbative expansion and for all observables.

In this work, we focused on the three observables that fall into the different categories. Linear impulse and angular impulse lie in the conservative sector. However, unlike the linear momentum operator, the angular momentum operator does not admit asymptotic states as its eigenstates. On the other hand, radiation measures the loss of matter momenta, which are radiated away to null infinity. The qualitative differences in the observables make this problem even more interesting to address. This is because not every observable can admit the cancellation of superclassical terms. The subtler properties of each of the observables are necessary for the cancellation of susperclassical terms. For example, Lorentz covariance of the angular momentum is needed to obtain Eqn.(\ref{eq:delLL}) and cancellation to occur at leading order. Similarly, the appearance of static modes in Eqn.(\ref{eq:N2to3}) in the classical limit of the Radiative field is crucial for the superclassical terms to vanish for non-zero momenta. Abstracting from these examples and from further down the proof, it will be interesting to classify the most general class of observables that admit smooth classical limit.

In inelastic channels, additional cancellation of superclassical terms occurs at the level of amplitude itself, ensuring that all superclassical terms that arise due to inelasticity cancel at the level of $N$ itself. More precisely, when the classical limit is taken for the tree-level amplitude, we show that the only non-zero contribution to singular terms in amplitude come from static mode photon contributions, as can be seen from Eqn.(\ref{eq:h-soft-lim}) and Eqn.(\ref{eq:S00}). However, upon integrating over inelastic photon channels in the final observable, they too vanish, as we find from analysing Eqn.(\ref{eq:A20-1ph}) and Eqn.(\ref{eq:pprod-stat-mod}). This results in a lack of any superclassical terms in the final observable. Interestingly, these cancellations for inelastic channels do not depend on the properties of observables and only depend on the interactions. In this paper, we showed it for the scalar QED but the proof can be easily extended for gravity. The argument for elastic channels goes on directly for gravitons, or for any massless propagators, as long as the coupling constants scale in a similar fashion. In inelastic channels, we only need to consider the classical limit of additional tree-level diagrams for gravity that involve self-interacting gravitons to estimate the $\hbar$ scaling of $N$. Alternatively, we saw, from comparing Eqn.(\ref{eq:main-inel}) and Eqn.(\ref{eq:S00}), the classical limit in the KMOC formalism for the external Gauge-bosons corresponds to the soft limit for the tree-level inelastic scattering in $3+1$ dimensions. So, we may directly invoke generalized tree-level soft theorems, proved for complex scalar interactions with photons, gravitons, and gluons in \cite{c:Inf-set-sof-thm-gg-th-wti} to show the vanishing of superclassical terms for any radiating Gauge-Bosons. In fact, it would be interesting to find the most general class of theories and interactions involving massless radiations that admit a smooth classical limit by allowing similar intricate superclassical cancellations in the inelastic channels. This may depend on how the tree-level soft theorems proved in \cite{c:Inf-set-sof-thm-gg-th-wti} extend to the soft behavior of other massless bosons such as the Nambu-Goldstone modes. On the other hand, concerning the matter in the theory, the internal massive matter propagators seem to play a role in tree-level inelastic scattering, as can be seen from Fig. (\ref{fig:fey-n-phot}). However, in the classical limit, the results we prove in the paper are expected to hold for Fermionic matters directly since the numerator of Fermionic propagators in the classical limit may not contain any outstanding terms.

In a different direction, it may also be worth investigating the role of gravitational background in the well-defined behavior of the classical limit. The smoothness of the classical limit in the KMOC formalism we proved in this paper corresponds to the usual QFT in Minkowski background. We saw through the paper that the Poincaré symmetry has been essential for intricate cancellation of superclassical terms, employing both Lorentz covariance of observables and energy-momentum conservation in the scattering to achieve the smoothness of the classical limit. Since the KMOC formalism has also been extended to curved backgrounds \cite{c:class-phy-curv-back}, it will be very interesting to study if these problems manifest when the classical limit of the physical observables is taken in the presence of curved backgrounds and how the problems may be handled. More precisely, what ingredients are essential for defining a physical observable with a well-defined classical limit in the presence of a background gravity. Or vice versa, which quantum theories are physical and consistent with the smooth classical limit.

Finally, we discuss the application of exponential representation within the KMOC formalism. As we showed in the paper, the formalism allows us to organize the terms very efficiently. The formalism gives us compact, elegant, and recursive formulas for classical observables incorporating all conservative and radiative contributions. We used the expansion, such as in Eqn.(\ref{eq:chan}), to organize the classical contributions on the basis of channels. We proved that the classical limits are those with elastic channel or just single-photon insertions between $N$ and $A_n^O$. Any higher number of photon insertions between them is necessarily quantum as we found from Eqn.(\ref{eq:h-fac-inel}). So, we have isolated the set of terms that contribute to the classical limit. For the elastic channel, explicit recursive expressions in $N$ have been provided for the classical observables, which remain after the cancellation of superclassical terms. This includes Eqn.(\ref{eq:ang.ansz.finr}) for the angular impulse and Eqn.(\ref{eq:class-rad}) for the Radiative field. They are analogs of the elastic channel recursive relation for linear impulse of Eqn.(\ref{eq:elas-chan-10}), obtained in \cite{exp-S-Mat-KMOC}. For inelastic channel contribution to the classical observables, we only need to consider diagrams that correspond to single-photon insertions between $N$ and $A_n^O$ i.e. $k=1$ in Eqn.(\ref{eq:chan}). The absence of superclassical terms allows us to directly pick out the terms from $N$ which are classically relevant i.e. those scaling with $\hbar^0$ in the final observable and ignore others such as the static modes from the inelastic tree-level diagrams. An interesting direction of work would be to directly evaluate these expressions to obtain the classical results. Since inelastic cancellations are independent of observables and depend only on interaction details, it may be a worthwhile task to pin down the terms in the inelastic scattering amplitude, or more precisely the corresponding $N$, which contributes classically by perhaps restricting to a certain gauge theory or gravity without worrying about the observables.

\section*{Acknowledgements}
We are thankful to Alok Laddha for suggesting the problem, many helpful discussions, and also helping with the draft. We are thankful to Amit Suthar, Samim Akhtar, and Manu Akavoor for the discussions. We also thank CSM (Chennai Strings Meet) 2024 at IMSc Chennai for allowing me to present this work.

\appendix

\section{Elastic Channel Contributions}
\label{ap:el-ch-cny}

We consider the observable defined by Eqn.(\ref{eq:O(p1p2)p1p2}). 
We isolate the terms that contain only elastic channels in Eqn.(\ref{eq:chan}) by inserting $2$-particle states between $ N $ and $ A_{n-1}^O $. These terms correspond to $k = 0$ in Eqn.(\ref{eq:chan}). Thus, in this section, we prove the following:
\framedtext{
\begin{align}
    \frac{1}{\hbar} & \int d \phi(\tl{p}_1) d \phi(\tl{p}_2) ( \la p_1', p_2' | N|\tl{p}_1, \tl{p}_2\ra \,  \la \tl{p}_1, \tl{p}_2 | A_{n-1}^O| p_1, p_2 \ra  
   \nonumber\\
   & - \la p_1', p_2' | A_{n-1}^O | \tl{p}_1, \tl{p}_2 \ra \, \la \tl{p}_1, \tl{p}_2 | N | p_1, p_2 \ra )\sim \frac{1}{\hbar^2} \, \hat{\del}^{(4)} (p_1'+p_2'-p_1-p_2).
 \label{eq:2to2ins}
\end{align}
}
\\
We factorize the delta function in the L.H.S. of Eqn.(\ref{eq:2to2ins}) as follows:
\begin{align}
\la p_1', p_2' & | N | p_1, p_2 \ra = N(p_1, p_2 \to p_1+q, p_2-q) \, \hat{\delta}^{(4)}(p_1' + p_2 ' - p_1 - p_2) 
\nonumber\\
& \text{and}\quad \la p_1', p_2' | A_n^O | p_1, p_2 \ra = A_n^O (p_1, p_2, q) \, \hat{\delta}^{(4)}(p_1' + p_2 ' - p_1 - p_2).
\end{align}
In $N (p_1, p_2  \to p_1', p_2')$, the momenta are on-shell, satisfying $2 p_1 \cdot q + q^2 = 0$ and $ 2 p_2 \cdot q - q^2 = 0$. Therefore, we can represent $N (p_1, p_2  \to p_1', p_2')$ as $N(p_1\cdot p_2, q^2)$ in a Lorentz covariant manner. Define $ k_1 = \tl{p}_1 - p_1$ and $ k_2 = \tl{p}_2 - p_2$ in Eqn.(\ref{eq:2to2ins}). Thus, the L.H.S. of Eqn.(\ref{eq:2to2ins}) becomes:
\begin{align}
 A_n^O  (  p_1, p_2, q) = & \hbar^3  \int \hat{d}^4 \ovk  \: \hat{\del}((p_1 + \hbar \ovk)^2 - m_1^2) \hat{\del}((p_2 -\hbar \ovk)^2 - m_2^2)
 \nonumber \\
   & \times  (N (p_1 \cdot p_2 + \hbar \ovk \cdot (p_2 - p_1) - \hbar^2 k^2, \hbar^2(\ovq-\ovk)^2) A_{n-1}^O ( p_1, p_2, \hbar  \ovk ) 
\nonumber\\
 & - A_{n-1}^O (p_1+ \hbar \ovk, p_2- \hbar \ovk, \hbar(\ovq - \ovk)) \, N ( p_1 \cdot p_2, \hbar^2 \, \ovk^2 )).
\label{eq:elas-chan-4}
\end{align} 

We separate the first and second terms and label them $f_O$ and $f_O^\prime$, respectively. The first term is given by:
\begin{align}
f_O = \hbar \int & \hat{d}^4 \ovk \: \hat{\del}(2p_1 \cdot \ovk + \hbar \ovk^2) \: \hat{\del}(2p_2 \cdot \ovk - \hbar \ovk^2) \:  N (p_1 \cdot p_2, \hbar^2(\ovq-\ovk)^2) 
A_{n-1}^O ( p_1, p_2, \hbar  \ovk ).
\label{eq:elas-chan-5}
\end{align} 
We used the on-shell condition: $2p_1 \cdot \ovk + \hbar \ovk^2 = 2p_2 \cdot \ovk - \hbar \ovk^2 = 0$. 
Changing the integration variable $\ovk $ to $ - \ovk + \ovq $ in the second term and applying the on-shell conditions for $p_1'$ and $p_2'$, we obtain:
\begin{align}
f^{\prime}_O = \hbar \int \hat{d}^4 \ovk \: & \hat{\del}(2p_1 \cdot \ovk + \hbar \ovk^2 + 2 \hbar \ovk.(\ovq - \ovk)) \: \hat{\del}(2p_2 \cdot \ovk - \hbar \ovk^2 - 2 \hbar \ovk.(\ovq - \ovk)) \:  
\nonumber\\
& \times N (p_1 \cdot p_2, \hbar^2 (\ovq - \ovk)^2)  A_{n-1}^O ( p_1 + \hbar (\ovq-\ovk), p_2 - \hbar (\ovq-\ovk), \hbar  \ovk ).
\label{eq:elas-chan-6}
\end{align} 
We perform a Taylor expansion of the terms in Eqn.(\ref{eq:elas-chan-6}) as follows:
\begin{align}
 &\hspace*{-5.3in} \hat{\del}(2p_1 \cdot \ovk + \hbar \ovk^2 + 2 \hbar \ovk.(\ovq - \ovk)) = \hat{\del}(2p_1 \cdot \ovk +  \hbar \ovk^2) + \hbar (\ovq - \ovk)^\mu \nabla_\mu \hat{\del}(2p_1 \cdot \ovk +  \hbar \ovk^2),
\nonumber\\ 
&\hspace*{-5.3in} \hat{\del}(2p_2 \cdot \ovk - \hbar \ovk^2 - 2 \hbar \ovk.(\ovq - \ovk)) =  \hat{\del}(2p_2 \cdot \ovk - \hbar \ovk^2) + \hbar (\ovq - \ovk)^\mu \nabla_\mu \hat{\del}(2p_2 \cdot \ovk -  \hbar \ovk^2),
\nonumber\\
A_{n-1}^O ( p_1 + \hbar (\ovq-\ovk), p_2 - \hbar (\ovq-\ovk), \hbar  \ovk )
= A_{n-1}^O ( p_1 , p_2, \hbar  \ovk ) + \hbar (\ovq - \ovk)^\mu \nabla_\mu A_{n-1}^O ( p_1 , p_2, \hbar  \ovk ),
\label{eq:expns}
\end{align}
where $ \nabla_\mu$ is given by Eqn.(\ref{eq:nabla}). Substituting Eqn.(\ref{eq:expns}) into Eqn.(\ref{eq:elas-chan-6}), we find that the leading order terms in Eqn.(\ref{eq:elas-chan-6}) as $\hbar \to 0$ completely cancel the terms in Eqn.(\ref{eq:elas-chan-5}).
Hence, $f_O- f^{\prime}_O = 0$ at leading order in $\hbar$. In the expansion Eqn.(\ref{eq:expns}), we have explicitly retained the $\hbar$ correction within the delta functions. This ensures the cancellation of terms in the sub-leading order, which does not contribute classically. In the classical limit, only the derivative correction term survives, contributing an extra factor of $\hbar$. Thus, Eqn.(\ref{eq:elas-chan-4}) simplifies to:
\begin{align}
A_n^O ( p_1, p_2, q)&  = 
\nonumber\\
  - & \hbar^2 \int \hat{d}^4 \ovk\:\:N ( p_1 \cdot p_2, \hbar^2 \ovk^2 )
(\ovq - \ovk)^\mu \nabla_\mu (\hat{\del}(2p_1 \cdot \ovk ) \: \hat{\del}(2p_2 \cdot \ovk ) \: A_{n-1}^O (p_1, p_2, \hbar(\ovq - \ovk))). 
\label{eq:elas-chan-10}
\end{align} 
Since $N(p_1, p_2  \to p_1', p_2') \sim 1/\hbar^2$ and $ A_{n-1}^O ( p_1, p_2, q) \sim 1/\hbar^2$, it follows that $ A_{n}^O ( p_1, p_2, q) \sim 1/\hbar^2$. Hence, the cancellation of superclassical terms is established for the elastic channel.

\section{Identity}
\label{ap:imp-id}
We give a proof for the following identity: 
\framedtext{\begin{align}
\sum_\pi \prod_{j=1}^{k}  \frac{1}{ p_1 . ( \sum _{l = 1}^{j}\ovr_{i_{\pi(l)}} )} = \prod_{l = 1}^{k}\frac{1}{ p_1 \cdot \ovr_{i_{l}}}.
\label{eq:prop-uni}
\end{align}}
We prove this by induction. For the base case $k=2$: 
\begin{align}
\left(\frac{1}{ p_1 .\ovr_{i_1}}  \frac{1}{ p_1 .(\ovr_{i_1}+\ovr_{i_2})}+ \frac{1}{p_1 .\ovr_{i_2}}  \frac{1}{ p_1 .(\ovr_{i_2}+\ovr_{i_1})}\right) 
= \left(\frac{1}{ p_1 .\ovr_{i_1}}  \frac{1}{ p_1 .\ovr_{i_2}} \right).
\end{align}
 We assume that this hypothesis holds for $k = n-1$. Consider L.H.S. of Eqn.(\ref{eq:prop-uni}) for $k=n$. Since the factor 
\begin{align}
\frac{1}{ p_1 \cdot ( \sum _{l = 1}^{n}\ovr_{i_{\pi(l)}} )} 
\end{align}
is invariant under any permutation $\pi \in S_n$, we can factor it out and write the L.H.S. as:
\begin{align}
\text{L.H.S.} =  \frac{1}{ p_1 \cdot ( \sum _{l = 1}^{ n}\ovr_{i_l})}  \sum_\pi \prod_{j=1}^{n-1} \frac{1}{ p_1 \cdot ( \sum _{l = 1}^{j}\ovr_{i_{\pi(l)}})}, 
\label{eq:lhs-mult}
\end{align}
where the sum runs over all permutations $\pi$ of the $n$ indices. If $\pi (n) = m$, then $\ovr_{i_m}$ is excluded from the inner sum over $l$ for all $j<n$ in Eqn.(\ref{eq:lhs-mult}). Thus, we can rewrite the expressions as:
\begin{align}
 \text{L.H.S.} = \frac{1}{  p_1 \cdot ( \sum _{l = 1}^{n}\ovr_{i_l})}  \sum_{\pi(n)=1}^{\pi(n)=n} \sum_{\sigma \in S_{n-1}} \prod_{1 \leq j \leq n}^{j \neq \pi(n)} \frac{1}{ p_1 \cdot ( \sum_{1 \leq l \leq j}^{l \neq \pi(n)}\ovr_{i_{\sig(l)}})}, 
\end{align}
where $\sig$ permutes the $n-1$ elements of the set $\{1,2 \ldots n\} \cap \{\pi(n)\}^c$.
Substituting Eqn.(\ref{eq:prop-uni}) for $k = n-1$, the L.H.S. becomes:
\begin{align}
  \frac{1}{ p_1 \cdot ( \sum _{l = 1}^{n}\ovr_{i_l})}  \sum_{\pi(n)=1}^{\pi(n)=n}  \prod_{1 \leq l \leq n}^{l \neq \pi(n)} \frac{1}{p_1 \cdot \ovr_{i_{l}} }  
= 
 \frac{1}{  p_1 \cdot ( \sum _{l = 1}^{n}\ovr_{i_l})} \frac{p_1 \cdot ( \sum _{l = 1}^{n}\ovr_{i_l})}{ \prod_{1 \leq l \leq n}  p_1 \cdot \ovr_{i_{l}} }, 
\end{align}
where we have multiplied and divided each term in the sum by $ p_1 \cdot \ovr_{i_{\pi(n)}} $. Thus, this is the same as the R.H.S. of Eqn.(\ref{eq:prop-uni}), establishing the identity:
\begin{align}
\sum_\pi \prod_{j=1}^{n} \frac{1}{ p_1 \cdot ( \sum _{l = 1}^{j}\ovr_{i_{\pi(l)}} )} = \prod_{l = 1}^{n} \frac{1}{p_1 \cdot \ovr_{i_{l}}}.
\label{eq:perm-nop-1}
\end{align} 

\section{Static Mode Computation for $A_2^O$}
\label{A:A2O-1ph}

In this section, we evaluate the expression for the static-mode contribution to $A_2^O$ with a single-photon insertion. In particular, we evaluate the following:
\begin{align}
 \la p_1', p_2' | A_2^O | p_1, p_2 \ra = \frac{1}{\hbar}\int d \phi(\tl{p}_1, \tl{p}_2, & \Upsilon_1^{\vec{r},\vec{\sig}}) (\la p_1', p_2' | N| \tl{p}_1, \tl{p}_2,\Upsilon_1^{\vec{r},\vec{\sig}}\ra \, \la \tl{p}_1, \tl{p}_2,\Upsilon_1^{\vec{r},\vec{\sig}} | A_1^O| p_1, p_2 \ra  
\nonumber\\
- & \la p_1', p_2' | A_1^O | \tl{p}_1, \tl{p}_2,\Upsilon_1^{\vec{r},\vec{\sig}} \ra \, \la \tl{p}_1, \tl{p}_2,\Upsilon_1^{\vec{r},\vec{\sig}} | N | p_1, p_2 \ra ),
\label{eq:C111}
\end{align}
where $p_1' = p_1+q$, $p_2' = p_2 - q$, $\tl{p}_1 = p_1 + q_+$ and $\tl{p}_2 = p_2 -q_+ - r$.
Factorizing the delta function, we obtain:
\begin{align}
A_2^O(p_1, p_2 \to & p_1', p_2') = 
\nonumber\\
\frac{1}{\hbar}\sum_\sig \int \hat{d}^4 q_+ \: & \frac{\hat{d}^3 r}{2 |\vec{r} \, |} \: \hat{\del}^{(+)}(2 p_1 \cdot q_+ + q_+^2) 
 \, \hat{\del}^{(+)}(2 p_2 \cdot (r+q_+) - (r+q_+)^2) \, (N(\tl{p}_1,\tl{p}_2, r, \sig \to p_1', p_2')
\nonumber\\
 & \times A_1^O (p_1, p_2 \to \tl{p}_1, \tl{p}_2, r, \sig) - A_1^O(\tl{p}_1, \tl{p}_2, r, \sig \to p_1', p_2') \, N(p_1, p_2 \to \tl{p}_1, \tl{p}_2, r, \sig)).
\label{eq:C22}
\end{align}
We use $\la l | N |  k \ra^* = \la k | N |  l \ra$ and $\la l | A_1^O |  k \ra^* = - \la k | A_1^O |  l \ra $ to exchange the incoming and outgoing states. Finally, substituting Eqn.(\ref{eq:A1phot}) for $k=1$ into Eqn.(\ref{eq:C22}) and taking the classical limit, we arrive at:
\begin{align}
A_2^O&(p_1, p_2 \to p_1', p_2') = 
 -\hbar^3 \sum_\sig \int \hat{d}^4 \ovq_+ \: \frac{\hat{d}^3 \ovr}{2 |\vec{\ovr} \, |} \hat{\del}^{(+)}(2 p_1\cdot \ovq_+) \, \hat{\del}^{(+)}(2 p_2 \cdot (\ovr+\ovq_+))
\nonumber\\
&N^*( p_1', p_2' \to \tl{p}_1, \tl{p}_2, r, \sig) \, N ( p_1, p_2 \to \tl{p}_1, \tl{p}_2, r, \sig) 
\left(  (2 \ovq_+^\mu  - \ovq^\mu) \nabla_\mu -  2 \ovr^\mu  \dodo{}{p_2^\mu} \right) O(p_1,p_2) .
\label{eq:A2}
\end{align}
Unlike in the case of elastic channel, the leading order term of $N A_1^O$  does not cancel the corresponding leading order term of $A_1^O N$ for the inelastic channel in Eqn.(\ref{eq:C111}).
For the tree level in $N$, the leading order term in $\hbar$ is given by: 
\begin{align}
N(p_1, p_2 \to p_1', p_2', r, \sigma) = i \frac{8 \, \bar{e}^3 \, p_1 \cdot p_2}{\hbar^{7/2}} \left(\frac{p_1 \cdot \eps_p(r)}{(\bar{q}+\ovr)^2+i \eps} \, \hat{\del}(2 p_1 \cdot \ovr) +  \frac{p_2 \cdot \eps_p(r)}{\bar{q}^2+i \eps} \, \hat{\del}(2 p_2 \cdot \ovr)\right),
\label{eq:N2to31}
\end{align}
where $p_1' = p_1 + q$ and $p_2' = p_2 - q - r$.
Substituting the tree-level contribution into Eqn.(\ref{eq:A2}) and using Eqn.(\ref{eq:msure}), we obtain:
\begin{align}
A_2^O(p_1, p_2 \to p_1', p_2') = &
\nonumber\\
-\frac{8 \, \bar{e}^6 \, (Q_1 \, Q_2 \, p_1 \cdot p_2)^2 } {\hbar^4} & \sum_p \int \hat{d}^4 \ovq_+ \:|\vec{\ovr}| \: \hat{d} |\vec{\ovr}|\sin \tht \: \hat{d} \tht \: \hat{d} \phi \: \hat{\del}^{(+)}(2 p_1\cdot \ovq_+) \,\hat{\del}^{(+)}(2 p_2 \cdot (\ovr+\ovq_+))
\nonumber\\
&\times \left(\frac{Q_1 \, p_1 \cdot \eps^*_p(r)}{(\ovq_++\ovr - \ovq)^2-i \eps} \frac{\hat{\del}(|\vec{\ovr}|)}{(p_1^0- \vec{p}_1 \cdot \hat{r} )} +  \frac{Q_2 \, p_2 \cdot \eps^*_p(r)}{(\ovq_+ - \ovq)^2-i \eps} \frac{\hat{\del}(|\vec{\ovr}|)}{(p_2^0- \vec{p}_2 \cdot \hat{r} )} \right) 
\nonumber\\
& \times \left(\frac{Q_1 \, p_1 \cdot \eps_p(r)}{(\ovq_++\ovr)^2+i \eps} \frac{\hat{\del}(|\vec{\ovr}|)}{(p_1^0- \vec{p}_1 \cdot \hat{r} )} +  \frac{Q_2 \, p_2 \cdot \eps_p(r)}{\ovq_+^2+i \eps} \frac{\hat{\del}(|\vec{\ovr}|)}{(p_2^0- \vec{p}_2 \cdot \hat{r} )} \right)
 \nonumber\\
 & \times \left( (2 \ovq_+^\mu  - \ovq^\mu) \nabla_\mu -  2 \ovr^\mu  \dodo{}{p_2^\mu} \right) O(p_1,p_2),
\end{align}
where $p_1' = p_1+q$, $p_2' = p_2 - q$.

\section{Ansatz for $A_n^{L_1}$ for arbitrary channel}
\label{a:generic-anl1}

In this section, we show that a valid ansatz for $\la p_1', p_2' | A_n^O| p_1, p_2 \ra$ can be expressed as:
\framedtext{
\begin{align}
\la p_1', p_2' | A_n^{L_1} | p_1, p_2 \ra =  A_n^{L_1} (p_1, p_2, q_1) \, \hat{\delta}^{(4)}(p_1'+ p_2' - p_1 - p_2) 
\nonumber\\
&\hspace*{-2in} + A_n^{L_1} (p_1, p_2, q_1, q_2) \wedge \dodo{}{p_1}\hat{\delta}^{(4)}(p_1'+ p_2' - p_1 - p_2),
\end{align}
where $p_1' = p_1 + q_1$ and $p_2' = p_2 + q_2$.
}
To show this, we consider a generic term in $\la p_1', p_2' | A_n^O| p_1, p_2 \ra$:
\begin{align}
\la p_1', p_2' | N^m \hat{p}_{1} \wedge \dodo{}{\hat{p}_{1}} N^{n-m} | p_1, p_2 \ra.
\end{align}
We evaluate this expression by inserting a complete set of states as follows:
\begin{align}
\sum_{k\geq 0}\int d \phi(\tl{p}_1, \tl{p}_2, \Upsilon_k^{\vec{r},\vec{\sig}})\, \la p_1', p_2' | N^m \hat{p}_{1} \wedge \dodo{}{\hat{p}_{1}} | \tl{p}_1, \tl{p}_2, \Upsilon_k^{\vec{r},\vec{\sig}} \ra \,
 \la \tl{p}_1, \tl{p}_2 , \Upsilon_k^{\vec{r},\vec{\sig}}| N^{n-m} | p_1, p_2 \ra.
 \label{eq:def-D2}
 \end{align}
 We factorize the matrix elements as follows:
\begin{align}
 \la p_1', p_2' | N^m |\tl{p}_1, \tl{p}_2, \Upsilon_k^{\vec{r},\vec{\sig}} \ra = & \, N_m (p_1', p_2',\tl{p}_1, \tl{p}_2, \Upsilon_k^{\vec{r},\vec{\sig}}) \, \hat{\delta}^{(4)}(p_1'+ p_2' - \tl{p}_1 - \tl{p}_2 - \sum_{i=1}^{k} r_i),
\nonumber\\
\la \tl{p}_1, \tl{p}_2 , \Upsilon_k^{\vec{r},\vec{\sig}}| N^{n-m} | p_1, p_2 \ra = & \, N^*_{n-m} (p_1,p_2,\tl{p}_1, \tl{p}_2, \Upsilon_k^{\vec{r},\vec{\sig}}) \, \hat{\delta}^{(4)}(\tl{p}_1+ \tl{p}_2 + \sum_{i=1}^{k} r_i - p_1 - p_2).
\label{eq:N-fact}
\end{align}
For $N^{n-m}$, we used the Hermicity of $N$.
Substituting Eqn.(\ref{eq:N-fact}) into Eqn.(\ref{eq:def-D2}), we obtain:
\begin{align}
\sum_{k\geq 0}  \int d \phi &(\tl{p}_1, \tl{p}_2, \Upsilon_k^{\vec{r},\vec{\sig}})N^*_{n-m} (p_1,p_2,\tl{p}_1, \tl{p}_2, \Upsilon_k^{\vec{r},\vec{\sig}}) \hat{\delta}^{(4)}(\tl{p}_1+ \tl{p}_2 + \sum_{i=1}^{k} r_i - p_1 - p_2)
\nonumber\\
 & ( \, \tl{p}_1 \wedge \dodo{}{\tl{p}_1}(N_m (p_1', p_2',\tl{p}_1, \tl{p}_2, \Upsilon_k^{\vec{r},\vec{\sig}})) \, \hat{\delta}^{(4)}(p_1'+ p_2' - \tl{p}_1 - \tl{p}_2 - \sum_{i=1}^{k} r_i) 
\nonumber\\
& + N_m (p_1', p_2',\tl{p}_1, \tl{p}_2, \Upsilon_k^{\vec{r},\vec{\sig}}) \tl{p}_1 \wedge \dodo{}{\tl{p}_1}\hat{\delta}^{(4)}(p_1'+ p_2' - \tl{p}_1 - \tl{p}_2 - \sum_{i=1}^{k} r_i) \,).
\end{align}
Upon evaluation, the final expression for $\la p_1', p_2' | N^m \hat{p}_{1} \wedge \dodo{}{\hat{p}_{1}} N^{n-m} | p_1, p_2 \ra $ can be written as:  
\begin{align}
\sum_{k\geq 0}\frac{1}{k!}\sum_{\vec{\sig}} \int d \phi (\tl{p}_1) \prod_{i=1}^{k}d \phi (r_i) \hat{\delta}(\tl{p}_2^2 - m_2^2)\,  \tl{p}_1 \wedge \dodo{}{\tl{p}_1}\, (\, N_m (p_1', p_2',\tl{p}_1, \tl{p}_2, \Upsilon_k^{\vec{r},\vec{\sig}})\, )
\nonumber\\
&\hspace*{-3.5in} \times N^*_{n-m} (p_1,p_2,\tl{p}_1, \tl{p}_2, \Upsilon_k^{\vec{r},\vec{\sig}}) \hat{\delta}^{(4)}(p_1'+ p_2' - p_1 - p_2)
\nonumber\\
&\hspace*{-4.5in} + \sum_{k\geq 0} \frac{1}{k!} \sum_{\vec{\sig}} \int d \phi (\tl{p}_1) \prod_{i=1}^{k}d \phi (r_i) \hat{\delta}(\tl{p}_2^2 - m_2^2) \, N_m (p_1', p_2',\tl{p}_1, \tl{p}_2, \Upsilon_k^{\vec{r},\vec{\sig}}) 
\nonumber\\
 &\hspace*{-3.5in} \times N^*_{n-m} (p_1,p_2,\tl{p}_1, \tl{p}_2, \Upsilon_k^{\vec{r},\vec{\sig}})  \tl{p}_1 \wedge \dodo{}{p_1}\hat{\delta}^{(4)}(p_1'+ p_2' - p_1 - p_2 ),
\end{align}
where $\tl{p}_2 = p_1+ p_2 + \sum_i r_i - \tl{p}_1$.
Thus, we can write:
\begin{align}
\la p_1', p_2' | A_n^{L_1} | p_1, p_2 \ra =  A_n^{L_1} (p_1, p_2, q_1) \, \hat{\delta}^{(4)}(p_1'+ p_2' - p_1 - p_2) 
\nonumber\\
&\hspace*{-2in} + A_n^{L_1} (p_1, p_2, q_1, q_2) \wedge \dodo{}{p_1}\hat{\delta}^{(4)}(p_1'+ p_2' - p_1 - p_2),
\end{align}
where $p_1' = p_1 + q_1$ and $p_2' = p_2 + q_2$.

\section{Computation for $A_n^{A_\mu(k)} (p_1, p_2 \to p_1', p_2')$}
\label{Ap:rad-elas}

In the section, we evaluate $A_n^O (p_1, p_2 \to p_1', p_2')$ for $O = \sqrt{2 E_k} \sum_r a_k^r \eps^r_\mu (k)$. Substituting the first term of Eqn.(\ref{eq:Aedel-1}) for $A_{n-1}$ into Eqn.(\ref{eq:elas-ch-rad}), we obtain:
\begin{align}
 \la p_1', p_2' | A_n^O  | p_1, p_2 \ra  = 
& \frac{1}{\hbar}\sum_r  \int  \hat{d}^4 q_+ \, \hat{d}^4 q_- \: \hat{\del}((p_1+ q_+)^2 - m_1^2) \, \hat{\del}((p_2+ q_-)^2 - m_2^2)
\nonumber\\
& \times ( N((p_1+q_+) \cdot (p_2+q_-), \, \hbar^2 (\ovq_1 - \ovq_+)^2) \, A_{n-1}^O(p_1, \, p_2, \, q_+, \, k , \, r) 
\nonumber\\
& \times \hat{\del}^{(4)}(p_1'+p_2' - p_1 - p_2 -q_+ - q_- ) \,
  \hat{\del}^{(4)}(q_+ + q_- + k ) 
  \nonumber\\
&  - N(p_1 \cdot p_2, \, \hbar^2 \ovq_+^2) \, A_{n-1}^O(p_1+q_+, \, p_2+q_-, \, q_1-q_+, \, k , \, r)  \, 
  \hat{\del}^{(4)}(q_++q_-) 
  \nonumber\\
&  \times \hat{\del}^{(4)}(p_1'+p_2'+ k - p_1-p_2 -q_+-q_- ) ) \, \eps^r_\mu (k),
\label{eq:elas-ch-rad-2}
\end{align}
where $ q_+ = \tl{p}_1 - p_1$ and $q_- = \tl{p}_2 - p_2 $. Upon integrating over $q_-$, we get an overall delta function $\hat{\del}^{(4)}(p_1'+p_2'+ k - p_1 - p_2)$, which enforces total 4-momentum conservation. Using this momentum conservation and the on-shell conditions, we find that $2 \tl{p}_1 \cdot \tl{p}_2 = 2 p_1 \cdot p_2 -2 k \cdot (p_1+p_2)$. 
  Thus, the first term in Eqn.(\ref{eq:elas-ch-rad-2}), labeled as $f_O$, becomes:
\begin{align}
f_O = \hbar\int & \hat{d}^4 \ovq_+ \: \hat{\del}(2p_1\cdot \ovq_+ + \hbar \ovq_+^2) \, \hat{\del}(2p_2 \cdot (\ovq_+ +\ovk) - \hbar (\ovq_++\ovk)^2)
\nonumber\\
&  \times  N(p_1 \cdot p_2 - \hbar \ovk \cdot (p_1+p_2), \hbar^2 (\ovq_1 - \ovq_+)^2) \, A_{n-1}^O(p_1, p_2, \hbar \ovq_+, \hbar \ovk, r).
 \label{eq:rad-1strm-2}
 \end{align}
Here, following Eqn.(\ref{eq:Aedel-1}), we have suppressed both the overall delta function and the polarization vector.
Similarly, evaluating the second term $f_O'$ in Eqn.(\ref{eq:elas-ch-rad-2}), while suppressing the same kinematic factors and changing the integration variable via $\ovq_+ \to \ovq_1 - \ovq_+$, we obtain:
\begin{align}
f_O' = \hbar\int \hat{d}^4 \ovq_+ \: \hat{\del}(2p_1 \cdot ( \ovq_+ - \ovq_1 ) - \hbar ( \ovq_+ - \ovq_1 )^2 ) \, \hat{\del}(2p_2 \cdot ( \ovq_+ - \ovq_1 ) + \hbar ( \ovq_+ - \ovq_1 )^2)
\nonumber\\
  &\hspace*{-4.5in} \times A_{n-1}^O ( p_1 +  \hbar ( \ovq_1 - \ovq_+) ,  p_2 - \hbar  ( \ovq_1 - \ovq_+),  \hbar \ovq_+ , \hbar \ovk   ,  r ) \: N(p_1 \cdot p_2, \hbar^2 (\ovq_1-\ovq_+)^2).
\label{eq:elas-ch-rad-3-chg-var}
\end{align}
Terms inside the first delta function can be rewritten as:
\begin{align}
(2p_1\cdot \ovq_+ + \hbar \ovq_+^2) - \underbrace{(2p_1 \cdot \ovq_1 + \hbar \ovq_1^2)}_{= 0 \:\:(\text{on-shell})}  + 2 \hbar \ovq_+ \cdot (\ovq_1 - \ovq_+ ).
\label{eq:expfrd}
\end{align}

Similarly, we rewrite the terms inside the second delta function and use the on-shell condition for $p_2' = p_2 - q_1 -k$ to get
\begin{align}
 2 p_2 \cdot (\ovq_++\ovk) - \hbar  (\ovq_++\ovk)^2  - 2 \hbar (\ovq_++\ovk ) \cdot (\ovq_1 - \ovq_+) \underbrace{ - 2 p_2 \cdot (\ovq_1 +\ovk)+ \hbar (\ovq_1 +\ovk)^2}_{= 0 \:\:(\text{on-shell})}. 
\label{eq:expscnd}
\end{align}
Using Eqn.(\ref{eq:expfrd}) and Eqn.(\ref{eq:expscnd}), we Taylor expand the delta functions in Eqn.(\ref{eq:elas-ch-rad-3-chg-var}):
\begin{align}
\hat{\del}(2p_1\cdot \ovq_+  + \hbar \ovq_+^2 + 2 \hbar \ovq_+ &\cdot (\ovq_1 - \ovq_+ )) 
= \hat{\del}(2p_1\cdot \ovq_+ + \hbar \ovq_+^2) + \hbar (\ovq_1 - \ovq_+ )^\mu\nabla_\mu \hat{\del}(2p_1\cdot \ovq_+ + \hbar \ovq_+^2),
\nonumber\\
\hat{\del}(2p_2 \cdot (\ovq_++\ovk)  - \hbar  (\ovq_++&\ovk)^2  - 2 \hbar (\ovq_++\ovk ) \cdot (\ovq_1 - \ovq_+) ) 
\nonumber\\
= \hat{\del}(2p_2 \cdot (\ovq_+&+\ovk) - \hbar  (\ovq_++\ovk)^2 ) 
+ \hbar (\ovq_1 - \ovq_+ )^\mu\nabla_\mu \hat{\del} (2p_2 \cdot (\ovq_++\ovk) - \hbar  (\ovq_++\ovk)^2),
\label{eq:rad-exp-1}
\end{align}
 where $\nabla_\mu$ is defined in Eqn.(\ref{eq:nabla}). Similarly, we expand $A_{n-1}^O$ in Eqn.(\ref{eq:elas-ch-rad-3-chg-var}): 
\begin{align}
A_{n-1}^O ( p_1 +&  \hbar ( \ovq_1 - \ovq_+) ,  p_2 - \hbar  ( \ovq_1 - \ovq_+),  \hbar \ovq_+ ,  \hbar \ovk   ,  r )  
\nonumber\\
& = A_{n-1}^O ( p_1 , p_2, \hbar  \ovq_+, \hbar \ovk,r ) 
 + \hbar (\ovq_1 - \ovq_+)^\mu \nabla_\mu A_{n-1}^O ( p_1 , p_2, \hbar  \ovq_+, \hbar \ovk,r ).
\label{eq:rad-exp}
\end{align}

Finally, we expand the $N$ function in Eqn.(\ref{eq:rad-1strm-2}):
\begin{align} 
N( p_1  \cdot p_2 - \hbar \ovk & \cdot (p_1  +p_2), \hbar^2 (\ovq_1 - \ovq_+)^2)
\nonumber\\
& = \,N(p_1 \cdot p_2, \hbar^2 (\ovq_1 - \ovq_+)^2) - \hbar \ovk^\mu\bigtriangleup_\mu N(p_1 \cdot p_2, \hbar^2 (\ovq_1 - \ovq_+)^2),
\nonumber\\
& \text{where}\quad \bigtriangleup_\mu = \dodo{}{p_1^\mu}+\dodo{}{p_2^\mu}.
\label{eq:corr-in-1}
\end{align}
We substitute Eqn.(\ref{eq:rad-exp-1}) and Eqn.(\ref{eq:rad-exp}) into Eqn.(\ref{eq:elas-ch-rad-3-chg-var}) and  take the classical limit $\hbar \to 0$. Similarly, substituting Eqn.(\ref{eq:corr-in-1}) into Eqn.(\ref{eq:rad-1strm-2}) and taking the classical limit, we find that the leading order term in Eqn.(\ref{eq:elas-ch-rad-3-chg-var}) completely cancels the leading order term in Eqn.(\ref{eq:rad-1strm-2}), yielding:
\begin{align}
f_O = f_O' = \hbar\int & \hat{d}^4 \ovq_+ \: \hat{\del}(2p_1\cdot \ovq_+ + \hbar \ovq_+^2) \, \hat{\del}(2p_2 \cdot (\ovq_+ +\ovk) - \hbar (\ovq_++\ovk)^2)
\nonumber\\
&  \times  N(p_1 \cdot p_2, \hbar^2 (\ovq_1 - \ovq_+)^2) \, A_{n-1}^O(p_1, p_2, \hbar \ovq_+, \hbar \ovk, r).
\end{align}
Thus, the sub-leading order in $\hbar$ in Eqn.(\ref{eq:elas-ch-rad-2}) constitutes the leading non-vanishing contribution, given by:
\begin{align}
 A_n^O ( p_1 , p_2, & \hbar  \ovq_1,  \hbar \ovk,r ) =  -\hbar^2 \int \hat{d}^4 \ovq_+ \,(\hat{\del}(2p_1\cdot \ovq_+) \hat{\del}(2p_2 \cdot (\ovq_++\ovk)) \, A_{n-1}^O(p_1, p_2, \hbar \ovq_+, \hbar \ovk , r) 
\nonumber\\
 &\times \ovk^\mu\bigtriangleup_\mu N(p_1 \cdot p_2, \hbar^2 (\ovq_1 - \ovq_+)^2)
 +(\ovq_1 - \ovq_+ )^\mu\nabla_\mu( \hat{\del}(2p_1\cdot \ovq_+) \hat{\del}(2p_2 \cdot (\ovq_++\ovk)) 
 \nonumber\\
 & \times A_{n-1}^O(p_1, p_2, \hbar \ovq_+, \hbar \ovk , r)) \, N(p_1 \cdot p_2, \hbar^2 (\ovq_1 - \ovq_+)^2)). 
\end{align}
This expression contributes to the classical radiation.

\end{document}